\documentclass[floatfix,footinbib,reprint,twocolumn,prb,aps,a4paper,preprintnumbers,amsmath,amssymb,longbibliography]{revtex4-1}

\usepackage{graphicx}
\usepackage{subfigure}
\usepackage{enumitem}  
\usepackage{bbold} 
\DeclareGraphicsExtensions{.png,.pdf}
\usepackage{epsfig}
\usepackage{bm}% bold math
\usepackage{color}
\usepackage{makeidx}
\usepackage[colorlinks=true,linkcolor=blue,citecolor=blue,urlcolor=blue]{hyperref}
\usepackage{siunitx}
\usepackage{xcolor}
\usepackage{soul}
\setstcolor{red}

\usepackage[bbgreekl]{mathbbol}
\DeclareSymbolFontAlphabet{\amsmathbb}{AMSb}%
\DeclareSymbolFontAlphabet{\mathbb}{AMSb}
\usepackage{amsfonts}
\usepackage{mathbbol}
\usepackage[utf8]{inputenc}
\usepackage[english]{babel}

\newcommand{\edu}[1]{\textbf{\color{red} [Eduardo: #1]}}
\begin{document}
\title{\bf Short period magnetization texture of B20-MnGe explained by thermally fluctuating local moments}
\author{Eduardo Mendive-Tapia,$^{1}$ Manuel dos Santos Dias,$^{2}$ Sergii Grytsiuk,$^{2}$ Julie B.\ Staunton,$^{3}$ Stefan Blügel,$^{2}$ Samir Lounis,$^{2,4}$}
\affiliation{$^{1}$Department of Computational Materials Design, Max-Planck-Institut für Eisenforschung, 40237 Düsseldorf, Germany}
\affiliation{$^{2}$Peter Grünberg Institut and Institute for Advanced Simulation, Forschungszentrum  Jülich  and  JARA,  52425  Jülich,  Germany}
\affiliation{$^{3}$Department of Physics, University of Warwick, Coventry CV4 7AL, United Kingdom}
\affiliation{$^4$Faculty of Physics, University of Duisburg-Essen, 47053 Duisburg, Germany}

%%%%%%%%%%%%%%%%%%%%%%%%%%%%%%%%%%%%%%%%%%%%%%%%%%%%%%%%%%%%%%%%%%%%%%%%%%%%%%%%%%%
\begin{abstract}
B20-type compounds, such as MnSi and FeGe, host helimagnetic and skyrmion phases at the mesoscale, which are canonically explained by the combination of ferromagnetic isotropic interactions with weaker chiral Dzyaloshinskii-Moriya ones. Mysteriously, MnGe evades this paradigm as it displays a noncollinear magnetic state at a much shorter nanometer scale.  Here we show that the length scale and volume-dependent magnetic properties of MnGe stem from purely isotropic exchange interactions, generally obtained in the paramagnetic state. Our approach is validated by comparing MnGe with the canonical B20-helimagnet FeGe. The free energy of MnGe is calculated, from which we show how triple-q magnetic states can stabilize by adding higher-order interactions.
\end{abstract}
%%%%%%%%%%%%%%%%%%%%%%%%%%%%%%%%%%%%%%%%%%%%%%%%%%%%%%%%%%%%%%%%%%%%%%%%%%%%%%%%%%%

\maketitle

\section{Introduction}

Atom-scale magnetic moments spontaneously order into magnetic structures of increasing complexity, ranging from simple collinear ferromagnetic and antiferromagnetic arrangements to magnetic skyrmions~\cite{NagaosaTokura,Fert2017} and complex noncollinear helimagnets~\cite{Mackintosh1,HughesNat}, with several recent discoveries~\cite{Takagi2018a,MnGeNat,Kurumaji2019,Spethmann2020,Kamber2020}.
These noncollinear magnetic structures can host unique physical phenomena, such as the topological Hall effect~\cite{Taguchi2001,Schulz2012}, and are being explored for practical uses, such as non-volatile memory~\cite{Fert2013}, caloric cooling~\cite{Matsunami1,PhysRevX.8.041035}, and neuromorphic computing~\cite{Song2020}.
Understanding and predicting the properties of complex magnetic phases has thus both fundamental appeal and huge potential for technological developments.

The family of B20-type compounds~\cite{doi:10.1002/adma.201603227,NagaosaTokura,Kanazawa_2016} is a canonical example of a rich magnetic phase diagram arising from the balance between the chiral Dzyaloshinskii-Moriya interaction (DMI)~\cite{DZYALOSHINSKY1958241,PhysRev.120.91} and ferromagnetic exchange interactions.
This results in a helimagnetic ground state and the appearance of skyrmions when a magnetic field is applied close to the N\'eel transition temperature $T_\mathrm{N}$~\cite{Muhlbauer915,NagaosaTokura,doi:10.1002/adma.201603227,Lebech_1989,PhysRevLett.107.127203,FeCoYu,PhysRevLett.102.037204,FeCoSiYu,BEILLE1983399,PhysRevB.72.224431}.
As the DMI is much weaker than the ferromagnetic exchange, the length scales of the helimagnetic phases are large compared to the distance between the atoms~\cite{Bak1980}, for example $\lambda\approx\SI{18}{\nano\meter}$~\cite{ISHIKAWA1976525,Muhlbauer915} and $\lambda\approx\SI{70}{\nano\meter}$~\cite{Lebech_1989} for MnSi and FeGe, respectively. 
This canonical picture mysteriously fails to explain MnGe, which instead of helimagnetism has a complex triple-q (3q) magnetic ground state with a much shorter period of $\lambda= 3-\SI{6}{\nano\meter}$~\cite{PhysRevLett.106.156603,MnGeNat}.
This 3q state stabilizes at all temperatures below $T_\mathrm{N}=\SI{180}{\kelvin}$, which leads to a very different temperature-magnetic field phase diagram in which skyrmions are absent~\cite{MnGeNat,PhysRevLett.106.156603,PhysRevB.88.064409,PhysRevB.86.134425,PhysRevB.96.220414,MnGeNatComm2016}.
Lastly, there is experimental evidence for strong thermal fluctuations in MnGe for a wide range of temperatures below $T_\mathrm{N}$~\cite{PhysRevB.90.144401,PhysRevB.93.174405}, which calls for their inclusion in any theoretical explanation of the magnetic properties.

Different experimental groups have pointed out that to understand MnGe within the canonical picture described above would require unrealistically high DMI values~\cite{MnGeNat,Altynbaev2016}.
Instead, competing long-ranged isotropic interactions akin to the famous Ruderman-Kittel-Kasuya-Yosida (RKKY) interactions~\cite{PhysRev.96.99,10.1143/PTP.16.45,PhysRev.106.893} can generate helimagnetic states with periods of a few nanometers, as found in the heavy rare earth elements~\cite{Mackintosh1,PhysRevLett.118.197202}, with the DMI playing a secondary role.
However, this explanation has not been demonstrated either experimentally or through first-principles calculations yet.
In fact, the zero-temperature first-principles studies carried out so far for MnGe~\cite{Gayles2015,Kikuchi2016,Koretsune2018,Mankovsky2018a,Bornemann_2019,SergiiNatComm} find the DMI to be weak and indicate that the isotropic pairwise interactions stabilize ferromagnetism, which falls back into the canonical picture of B20 magnetism and provides no explanation for the 3q state and its short period.

Here we demonstrate that the mechanism behind the magnetism of MnGe can be captured from first-principles by rigorously describing the thermal fluctuations of the local magnetic moments.
Pairwise interactions obtained in the paramagnetic limit are suitable quantities to robustly describe its instabilities, from which we explain the origin of the short period magnetism and the experimentally found volume-dependent magnetic properties of MnGe.
Our results are illustrated by a comparison with FeGe, whose helimagnetic state is underpinned by the DMI.
We also show that a 3q-state is easily stabilized if weak higher-order interactions are added to our pairwise interactions, and that the critical magnetic field beyond which the ferromagnetic state becomes stable is quantitatively reproduced.

The paper is organized as follows. In Sec.\ \ref{SecII} we introduce our first-principles theory of magnetism at finite temperature, which provides the Gibbs free energy of a magnetic material containing pairwise and higher order magnetic interactions. We explain how to Fourier transform the isotropic and DMI-like pairwise interactions to study the stability of wave-modulated magnetic states in the high paramagnetic temperature limit, and how to minimize the free energy to obtain the magnetic properties at lower temperatures. In Sec.\ \ref{SecIII} we show the application of our theory to MnGe and FeGe and study the potential effect of higher than pairwise interactions on the stability of triple-q states in MnGe. Finally, in Sec.\ \ref{SecConc} we present the conclusions of our work. Appendices \ref{Modulation} and \ref{1q} contain further information about our first-principles theory.

%\mdsd{Methods}

\section{First-principles theory of the Gibbs free energy}
\label{SecII}

We use a Disordered Local Moment (DLM) theory~\cite{0305-4608-15-6-018} to efficiently describe how the electronic structure responds to thermal fluctuations of the local moments.
This approach provides, for example, temperature-dependent magnetic anisotropy~\cite{PhysRevLett.93.257204,PhysRevB.74.144411,PhysRevLett.120.097202,PhysRevB.99.054415}, magnetic phase transitions and diagrams~\cite{PhysRevB.99.144424,PhysRevLett.118.197202,PhysRevB.95.184438,PhysRevMaterials.1.024411,PhysRevLett.115.207201}, and caloric effects~\cite{doi:10.1063/5.0003243}.
Our DLM theory is built upon the identification of robust local magnetic moments at certain atomic sites $\{n\}$, with orientations prescribed by unit vectors $\{\hat{\textbf{e}}_n\}$.
The main tenet of our approach is the assumption of a time-scale separation between the evolution of $\{\hat{\textbf{e}}_n\}$ in comparison to a rapidly adapting underlying electronic structure~\cite{0305-4608-15-6-018}.
Following this adiabatic approximation, density functional theory (DFT) calculations constrained to different magnetic configurations $\{\hat{\textbf{e}}_n\}$ can be performed to describe different states of magnetic order arising at finite temperatures.
The finite-temperature magnetic state of the material is then obtained by appropriate statistical averages over sets of local moment orientations.
These define the local order parameters
\begin{equation}
\left\{\textbf{m}_n \equiv \langle\hat{\textbf{e}}_n\rangle\right\},
\label{EqOP}
\end{equation}
which encode the temperature dependence~\cite{0305-4608-15-6-018,PhysRevB.99.144424}.

In Sec.\ \ref{IIA} we firstly describe our theory in the high temperature, paramagnetic, limit, in which we also explain how to Fourier transform the magnetic interactions in multi-sublattice systems for the study of wave-modulated magnetic states. Sec.\ \ref{Frame} presents a mean-field treatment to carry out the averages over the local moment orientations such that a Gibbs free energy of a magnetic material can be provided and studied at different temperatures. A method to minimize the free energy and obtain magnetic phase diagrams is shown in Sec.\ \ref{min}. In Sec.\ \ref{CompDetail} we give details of our computational method.

\subsection{High temperature paramagnetic limit}
\label{IIA}

At high enough temperatures the magnetic ordering of a magnetic material is described by fully disordered local moments whose orientations average to zero ($\{|\mathbf{m}_n| = 0\}$), i.e.\ the paramagnetic state.
We obtain the potential magnetic phases that stabilize upon lowering the temperature by studying the energetic cost of a small amount of magnetic order ($|\mathbf{m}_n| \ll 1$) introduced to the paramagnetic state.
A central outcome of our approach is a first-principles magnetic free energy $\mathcal{G}$~\cite{PhysRevB.99.144424}, which we expand here in terms of $\{\mathbf{m}_n\}$ in the spirit of the Ginzburg-Landau theory of phase transitions,
\begin{align}\label{eq:free_high_temp}
  \mathcal{G} &\approx -k_\mathrm{B}T\left(\ln 4\pi-\frac{3}{2} \sum_n |\mathbf{m}_n|^2\right) -\textbf{B}\cdot\sum\limits_{n}\mu_n\textbf{m}_n \nonumber\\
  &-\frac{1}{2}\sum\limits_{nn'}\left[ J_{nn'}^\mathrm{PM}\,\mathbf{m}_{n}\cdot\mathbf{m}_{n'}
  + \mathbf{D}_{nn'}^\mathrm{PM}\cdot \big(\mathbf{m}_{n} \times \mathbf{m}_{n'}\big)\right] \;.
\end{align}
The first term is the contribution from the magnetic entropy (see Sec.\ \ref{Frame}), with $k_\mathrm{B}$ the Boltzmann constant and $T$ the temperature, and the second term describes the effect of an applied magnetic field $\mathbf{B}$ on the local moments with sizes $\{\mu_n\}$. $J_{nn'}^\mathrm{PM}$ and $\mathbf{D}_{nn'}^\mathrm{PM}$ can be interpreted as mediating effective pairwise interactions, isotropic and DMI-like, respectively, calculated in the paramagnetic (PM) limit.

At zero temperature, the magnetic material is usually in a fully-ordered state ($|\mathbf{m}_n| = 1$), for instance a non-fluctuating ferromagnetic (FM) state.
In principle, we could then expand the energy by considering small transverse deviations of the order parameters and obtain zero-temperature pairwise interactions $J_{nn'}^\mathrm{FM}$ and $\mathbf{D}_{nn'}^\mathrm{FM}$.
We use a small deviation from the reference state in both the high-temperature (i.e.\ the paramagnetic state) and the zero-temperature limits in a perturbative method to evaluate these interactions directly from the respective electronic structures~\cite{PhysRevB.68.104436,PhysRevB.79.045209}, see Sec.\ \ref{CompDetail} for computational detail on this calculation.
Importantly, both sets of effective pairwise interactions are in principle different, which is a direct consequence of the presence of higher order magnetic interactions.

\subsubsection{Paramagnetic-helimagnetic phase transitions from high temperature paramagnetic interactions}
\label{Fourier}

The highest transition temperature $T_\text{max}$ from which the paramagnetic state is unstable to the formation of another magnetic phase can be calculated by studying the temperature dependence of $\mathcal{G}$ in Eq.\ (\ref{eq:free_high_temp}).
The process to obtain $T_\text{max}$ follows by firstly constructing the Hessian matrix of $\mathcal{G}$, $\zeta_{nn',\alpha\beta}=\frac{\partial^2\mathcal{G}}{\partial m_{n,\alpha}\partial {m}_{n',\beta}}$, where Greek letters are used to indicate spatial directions ($\{x,y,z\}$) and $m_{n,\alpha}$ is a spatial component of $\textbf{m}_n$. The next step is to find the condition that sets the determinant of the Hessian matrix zero at $\{\textbf{m}_n=\textbf{0}\}$,
%%%%%%%%%%%%%%%%%%%%%%%%%%%%%%%%%%%%%%%%%%%%%%%%%%%%%%%%%%%%%%%%%%%%%%%%%%%%%%%%%%%
\begin{equation}
\det\zeta\big\rvert_{\{\textbf{m}_n=\textbf{0}\}}=0.
\label{det0}
\end{equation}
%%%%%%%%%%%%%%%%%%%%%%%%%%%%%%%%%%%%%%%%%%%%%%%%%%%%%%%%%%%%%%%%%%%%%%%%%%%%%%%%%%%
Solving Eq.\ (\ref{det0}) provides a temperature at which a non-paramagnetic solution ($\textbf{m}_n\neq\textbf{0}$ at least in one site) becomes a minimum of $\mathcal{G}$.
Taking the second derivative of Eq.\ (\ref{eq:free_high_temp}) with respect to the order parameters shows that the components of $\zeta$ are
%%%%%%%%%%%%%%%%%%%%%%%%%%%%%%%%%%%%%%%%%%%%%%%%%%%%%%%%%%%%%%%%%%%%%%%%%%%%%%%%%%%
\begin{equation}
\zeta_{nn',\alpha\beta}\big\rvert_{\{\textbf{m}_n=\textbf{0}\}}=3k_\text{B}T\delta_{nn'}\delta_{\alpha\beta}-J^\text{PM}_{nn'}\delta_{\alpha\beta}-\sum_{\gamma}D^\text{PM}_{nn',\gamma}\epsilon_{\gamma\alpha\beta},
\label{EqHess0}
\end{equation}
%%%%%%%%%%%%%%%%%%%%%%%%%%%%%%%%%%%%%%%%%%%%%%%%%%%%%%%%%%%%%%%%%%%%%%%%%%%%%%%%%%%
%
where $\delta_{\alpha\beta}$ is the Kronecker delta and $\epsilon_{\gamma\alpha\beta}$ is the Levi-Civita symbol.
To solve Eq.\ (\ref{det0}) one needs to construct matrices of dimension as large as the long range nature of the magnetic interactions. An advantage of the paramagnetic state is that its crystal symmetry can be exploited and a lattice Fourier transform can be employed to study the stability of magnetic phases generally as functions of a wave vector $\textbf{q}$~\cite{PhysRevB.99.144424}. The primitive unit cell of the B20 compounds contains four magnetic sublattices. A lattice Fourier transform of the pairwise terms in Eq.\ (\ref{EqHess0}) that captures the potential complexity of these four degrees of freedom can be defined as
%%%%%%%%%%%%%%%%%%%%%%%%%%%%%%%%%%%%%%%%%%%%%%%%%%%%%%%%%%%%%%%%%%%%%%%%%%%%%%%%%%%
\begin{eqnarray}
\label{EqFT1}
& J^\text{PM}_{ss'}(\textbf{q})=\frac{1}{N_c}\sum\limits_{tt'}J^\text{PM}_{tst's'}\exp\left[-i\textbf{q}\cdot(\textbf{R}_{t}-\textbf{R}_{t'})\right], \\
& D^\text{PM}_{ss',\gamma}(\textbf{q})=\frac{1}{N_c}\sum\limits_{tt'}D^\text{PM}_{tst's',\gamma}\exp\left[-i\textbf{q}\cdot(\textbf{R}_{t}-\textbf{R}_{t'})\right],
\label{EqFT2}
\end{eqnarray}
%%%%%%%%%%%%%%%%%%%%%%%%%%%%%%%%%%%%%%%%%%%%%%%%%%%%%%%%%%%%%%%%%%%%%%%%%%%%%%%%%%%
where $N_c$ is the number of unit cells, and the site indices $n$ ($n'$) split into two different integers $t$ and $s$. $\{t,s\}$ decompose the position of a magnetic atom at site $n$, $\textbf{X}_n\equiv\textbf{X}_{ts}=\textbf{R}_t+\textbf{r}_s$, into a vector pointing at the origin of the corresponding unit cell, $\textbf{R}_t$, and a vector providing the relative position of the atom inside the cell, $\textbf{r}_s$.
The sums in Eqs.\ (\ref{EqFT1}) and (\ref{EqFT2}) are performed over $t$-indices only, i.e.\ over unit cell origins. For MnGe and FeGe the subindices $s$ and $s'$ label the four magnetic sublattices and so $J^\text{PM}_{ss'}(\textbf{q})$ and $D^\text{PM}_{ss',\gamma}(\textbf{q})$ are components of $4\times 4$ matrices, $J^\text{PM}(\textbf{q})$ and $D^\text{PM}_{\gamma}(\textbf{q})$ (where $\gamma=\{x,y,z\}$), respectively.

Applying the lattice Fourier transform defined above to Eq.\ (\ref{EqHess0}) gives
%%%%%%%%%%%%%%%%%%%%%%%%%%%%%%%%%%%%%%%%%%%%%%%%%%%%%%%%%%%%%%%%%%%%%%%%%%%%%%%%%%%
\begin{equation}
\zeta_{ss',\alpha\beta}(\textbf{q})=3k_\text{B}T\delta_{ss'}\delta_{\alpha\beta}-J^\text{PM}_{ss'}(\textbf{q})\delta_{\alpha\beta}-\sum_{\gamma}D^\text{PM}_{ss',\gamma}(\textbf{q})\epsilon_{\gamma\alpha\beta},
\label{EqHess}
\end{equation}
%%%%%%%%%%%%%%%%%%%%%%%%%%%%%%%%%%%%%%%%%%%%%%%%%%%%%%%%%%%%%%%%%%%%%%%%%%%%%%%%%%%
which allows to express Eq.\ (\ref{det0}) in a matrix form of tractable size,
%%%%%%%%%%%%%%%%%%%%%%%%%%%%%%%%%%%%%%%%%%%%%%%%%%%%%%%%%%%%%%%%%%%%%%%%%%%%%%%%%%%
\begin{equation}
\renewcommand{\arraystretch}{1.2}
\det\left[
3k_\mathrm{B} T
\mathcal{I}
-\mathcal{J}(\textbf{q})
\right]=0,
\label{EqHessM}
\end{equation}
%%%%%%%%%%%%%%%%%%%%%%%%%%%%%%%%%%%%%%%%%%%%%%%%%%%%%%%%%%%%%%%%%%%%%%%%%%%%%%%%%%%
where $\mathcal{I}$ is the identity matrix and
%%%%%%%%%%%%%%%%%%%%%%%%%%%%%%%%%%%%%%%%%%%%%%%%%%%%%%%%%%%%%%%%%%%%%%%%%%%%%%%%%%%
\begin{equation}
\renewcommand{\arraystretch}{1.2}
\mathcal{J}(\textbf{q})=
\begin{pmatrix}
 J^\text{PM}(\textbf{q})     &  D^\text{PM}_{z}(\textbf{q}) & -D^\text{PM}_{y}(\textbf{q}) \\
-D^\text{PM}_{z}(\textbf{q}) &  J^\text{PM}(\textbf{q})     &  D^\text{PM}_{x}(\textbf{q}) \\
 D^\text{PM}_{y}(\textbf{q}) & -D^\text{PM}_{x}(\textbf{q}) &  J^\text{PM}(\textbf{q})
\end{pmatrix}
\label{EqHessMb}
\end{equation}
%%%%%%%%%%%%%%%%%%%%%%%%%%%%%%%%%%%%%%%%%%%%%%%%%%%%%%%%%%%%%%%%%%%%%%%%%%%%%%%%%%%
is a 12$\times$12 matrix (four magnetic sublattices $\times$ three spatial dimensions) with components
%%%%%%%%%%%%%%%%%%%%%%%%%%%%%%%%%%%%%%%%%%%%%%%%%%%%%%%%%%%%%%%%%%%%%%%%%%%%%%%%%%%
\begin{equation}
\mathcal{J}_{ss'\alpha\beta}(\textbf{q})\equiv J^\text{PM}_{ss'}(\textbf{q})\delta_{\alpha\beta}+\sum_{\gamma}D^\text{PM}_{ss',\gamma}(\textbf{q})\epsilon_{\gamma\alpha\beta}.
\label{TotalMatrix}
\end{equation}
%%%%%%%%%%%%%%%%%%%%%%%%%%%%%%%%%%%%%%%%%%%%%%%%%%%%%%%%%%%%%%%%%%%%%%%%%%%%%%%%%%%

Eq.\ (\ref{EqHessM}) defines an eigenvalue problem for the diagonalization of $\mathcal{J}(\textbf{q})$,
%%%%%%%%%%%%%%%%%%%%%%%%%%%%%%%%%%%%%%%%%%%%%%%%%%%%%%%%%%%%%%%%%%%%%%%%%%%%%%%%%%
\begin{equation}
\sum_{s'\beta}\mathcal{J}_{ss'\alpha\beta}(\textbf{q})V_{s'\beta,p}(\textbf{q})=u_p(\textbf{q})V_{s\alpha,p}(\textbf{q}),
\label{EqM}
\end{equation}
%%%%%%%%%%%%%%%%%%%%%%%%%%%%%%%%%%%%%%%%%%%%%%%%%%%%%%%%%%%%%%%%%%%%%%%%%%%%%%%%%%%
whose eigenvalues, $\{u_p\}$, and eigenvectors, $\{V_{s\alpha,p}\}$, are functions of the wave vector $\textbf{q}$, and where $p=1,\dots,12$ enumerates the eigen-solution. Eqs.\ (\ref{EqHessM}) and (\ref{EqM}) show that the highest transition temperature as a function of the wave vector is given by the largest eigenvalue, i.e.\
\begin{equation}
\mathcal{J}_\text{max}(\textbf{q})\equiv \max[\{u_p(\textbf{q})\}]=u_{p_\text{max}}(\textbf{q}),
\label{EQJmax}
\end{equation}
 where $p_\text{max}$ is used to denote the index corresponding to this largest eigenvalue. $T_\text{max}$ is, therefore,
%%%%%%%%%%%%%%%%%%%%%%%%%%%%%%%%%%%%%%%%%%%%%%%%%%%%%%%%%%%%%%%%%%%%%%%%%%%%%%%%%%%
\begin{equation}
T_\text{max}=\frac{\mathcal{J}_\text{max}(\textbf{q}_{peak})}{3k_\text{B}}
,
\label{Eq3}
\end{equation}
%%%%%%%%%%%%%%%%%%%%%%%%%%%%%%%%%%%%%%%%%%%%%%%%%%%%%%%%%%%%%%%%%%%%%%%%%%%%%%%%%
where $\textbf{q}_{peak}$ is the wave vector at which the maximum value of $\mathcal{J}_\text{max}(\textbf{q})$ is found.
$\textbf{q}_{peak}$ describes the wave-modulation of the most stable magnetic state that the paramagnetic state is unstable to the formation of.
In particular, the effect of the DMI is removed by setting $D_{ss',\gamma}^\text{PM}=0$. This reduces the eigenvalue problem in Eq.\ (\ref{EqHessM}) to the diagonalization of the $4 \times 4$ matrix $J^\text{PM}(\textbf{q})$ only.
Fig.\ \ref{Fig1}(g,h) in Sec.\ \ref{SecIII} shows results obtained for the most stable solution as a function of $\textbf{q}$, $\mathcal{J}_\text{max}(\textbf{q})$, without DMI (diagonalization of $J^\text{PM}(\textbf{q})$ only, continuous lines) and with DMI (diagonalization of $\mathcal{J}(\textbf{q})$, dashed lines).
Results shown in Fig.\ \ref{Fig2} are obtained from the diagonalization of $J^\text{PM}(\textbf{q})$, i.e.\ without the DMI.

\subsection{Statistical mechanics of disordered local moments}
\label{Frame}

In our DLM theory, intermediate temperatures, $0\text{K}<T<T_N$, are described by intermediate values of the local order parameters $\{0<|\textbf{m}_n|<1\}$. The averages over the local moment orientations $\{\hat{\textbf{e}}_n\}$ defining them in Eq.\ (\ref{EqOP}) are obtained by considering a trial mean-field Hamiltonian consisting of a collection of internal magnetic fields at each magnetic site~\cite{0305-4608-15-6-018}
%%%%%%%%%%%%%%%%%%%%%%%%%%%%%%%%%%%%%%%%%%%%%%%%%%%%%%%%%%%%%%%%%%%%%%%%%%%%%%%%%%%
\begin{equation}
\mathcal{H}_0=-\sum_n\textbf{h}_n\cdot\hat{\textbf{e}}_n.
\label{EqH0}
\end{equation}
%%%%%%%%%%%%%%%%%%%%%%%%%%%%%%%%%%%%%%%%%%%%%%%%%%%%%%%%%%%%%%%%%%%%%%%%%%%%%%%%%%%
Eq.\ (\ref{EqH0}) defines a trial, single-site, probability distribution $P(\{\hat{\textbf{e}}_n\})=\prod\limits_n P_n(\hat{\textbf{e}}_n)=\frac{1}{Z_0}\prod\limits_n\exp\left[\beta\textbf{h}_n\cdot\hat{\textbf{e}}_n\right]$ with an associated partition function
$Z_0=\prod\limits_n\int d\hat{\textbf{e}}_n\exp\left[\beta\textbf{h}_n\cdot\hat{\textbf{e}}_n\right]=\prod\limits_n 4\pi\frac{\sinh\beta h_n}{\beta h_n}$,
where $\beta=1/k_\mathrm{B} T$ is the Boltzmann factor, and $h_n$ is the magnitude of the corresponding internal magnetic field. The magnetic order parameters are, therefore, given by the following expression (see Eq.\ (\ref{EqOP}))
%%%%%%%%%%%%%%%%%%%%%%%%%%%%%%%%%%%%%%%%%%%%%%%%%%%%%%%%%%%%%%%%%%%%%%%%%%%%%%%%%%%
\begin{equation}
\textbf{m}_n=\int d\hat{\textbf{e}}_n P_n(\hat{\textbf{e}}_n)\hat{\textbf{e}}_n=\left[-\frac{1}{\beta h_n}+\coth(\beta h_n)\right]\frac{\textbf{h}_n}{h_n}.
\label{Eqm}
\end{equation}
%%%%%%%%%%%%%%%%%%%%%%%%%%%%%%%%%%%%%%%%%%%%%%%%%%%%%%%%%%%%%%%%%%%%%%%%%%%%%%%%%%%
%
Owing to the single-site nature of Eq.\ (\ref{EqH0}) the magnetic entropy associated with the local moment orientations is a sum of single-site contributions,
%%%%%%%%%%%%%%%%%%%%%%%%%%%%%%%%%%%%%%%%%%%%%%%%%%%%%%%%%%%%%%%%%%%%%%%%%%%%%%%%%%%
\begin{equation}
%\mathcal{G}=-TS_{mag}-\textbf{B}\cdot\sum\limits_{n}\mu_n\textbf{m}_n
S_{mag}=\sum\limits_n S_n,
\label{EqSmagG}
\end{equation}
%%%%%%%%%%%%%%%%%%%%%%%%%%%%%%%%%%%%%%%%%%%%%%%%%%%%%%%%%%%%%%%%%%%%%%%%%%%%%%%%%%%
which are obtained by performing the single-site integrals
%%%%%%%%%%%%%%%%%%%%%%%%%%%%%%%%%%%%%%%%%%%%%%%%%%%%%%%%%%%%%%%%%%%%%%%%%%%%%%%%%%%
\begin{equation}
\begin{split}
S_n & =-k_\text{B}\langle\ln[P_n(\hat{\textbf{e}}_n)]\rangle=-k_\mathrm{B}\int d\hat{\textbf{e}}_n P_n(\hat{\textbf{e}}_n)\ln[P_n(\hat{\textbf{e}}_n)] \\
 & =k_\text{B}\left[1+\ln\left(4\pi\frac{\sinh(\beta h_n)}{\beta h_n}\right)-\beta h_n\coth(\beta h_n)\right].
\label{EqS1}
\end{split}
\end{equation}
%%%%%%%%%%%%%%%%%%%%%%%%%%%%%%%%%%%%%%%%%%%%%%%%%%%%%%%%%%%%%%%%%%%%%%%%%%%%%%%%%%%
%
Eq.\ (\ref{EqS1}) can be expanded in terms of $m_n=|\textbf{m}_n|$ using Eq.\ (\ref{Eqm}),
%%%%%%%%%%%%%%%%%%%%%%%%%%%%%%%%%%%%%%%%%%%%%%%%%%%%%%%%%%%%%%%%%%%%%%%%%%%%%%%%%%%
\begin{equation}
%\begin{split}
S_n=k_\mathrm{B}\Bigg(\ln 4\pi
 -\frac{3}{2}m_n^2
-\frac{9}{20}m_n^4
%-\frac{99}{350}m_n^6
%-\frac{1539}{7000}m_n^8
%-\frac{336}{1795}m_n^{10} \\
%& -\frac{846}{5083}m_n^{12}
%-\frac{313}{2074}m_n^{14}
%-\frac{215}{1562}m_n^{16}
%-\frac{476}{3803}m_n^{18}
%-\frac{157}{1399}m_n^{20}
-\dots\Bigg),
\label{EqS2}
%\end{split}
\end{equation}
%%%%%%%%%%%%%%%%%%%%%%%%%%%%%%%%%%%%%%%%%%%%%%%%%%%%%%%%%%%%%%%%%%%%%%%%%%%%%%%%%%%
which shows that the entropy in the paramagnetic limit is $S_{mag}=k_\mathrm{B}\sum\limits_{n}\left(\ln 4\pi-\frac{3}{2}m_n^2\right)$, as used in Eq.\ (\ref{eq:free_high_temp}).

\subsection{Minimization of the Gibbs free energy and magnetic phase diagrams}
\label{min}

The first-principles Gibbs free energy $\mathcal{G}$, shown in Eq.\ (\ref{eq:free_high_temp}) for the particular case of the paramagnetic limit, can be generally expressed at any value of $\{\textbf{m}_n\}$ as~\cite{PhysRevB.99.144424}
%%%%%%%%%%%%%%%%%%%%%%%%%%%%%%%%%%%%%%%%%%%%%%%%%%%%%%%%%%%%%%%%%%%%%%%%%%%%%%%%%%%
\begin{equation}
%\begin{split}
\label{EQG}
\mathcal{G}
= E(\{\textbf{m}_n\})
 -TS_{mag}(\{\textbf{m}_n\})
-\textbf{B}\cdot\sum\limits_{n}\mu_{n}\textbf{m}_n
,
%\end{split}
\end{equation}
%%%%%%%%%%%%%%%%%%%%%%%%%%%%%%%%%%%%%%%%%%%%%%%%%%%%%%%%%%%%%%%%%%%%%%%%%%%%%%%%%%%
where
%%%%%%%%%%%%%%%%%%%%%%%%%%%%%%%%%%%%%%%%%%%%%%%%%%%%%%%%%%%%%%%%%%%%%%%%%%%%%%%%%%%
\begin{equation}
\begin{split}
\label{EQE}
& E(\{\textbf{m}_n\}) = \\
& -\sum\limits_{nn',\alpha\beta}\left[J_{nn'}^\text{PM}\delta_{\alpha\beta}+\sum\limits_{\gamma}D_{nn',\gamma}^\text{PM}\epsilon_{\gamma\alpha\beta}\right]m_{n,\alpha}m_{n',\beta} +\text{h.o.}
,
\end{split}
\end{equation}
%%%%%%%%%%%%%%%%%%%%%%%%%%%%%%%%%%%%%%%%%%%%%%%%%%%%%%%%%%%%%%%%%%%%%%%%%%%%%%%%%%%
is an internal magnetic energy containing higher order than pairwise magnetic interactions, h.o., whose effect becomes negligible in the paramagnetic limit~\cite{PhysRevB.99.144424}.
The minimization of Eq.\ (\ref{EQG}) at different values of $T$ and $\textbf{B}$ provides the most stable magnetic states and corresponding magnetic phase diagrams as function of these quantities. Such a minimization can be accomplished by using the following expression for the internal magnetic field,
%%%%%%%%%%%%%%%%%%%%%%%%%%%%%%%%%%%%%%%%%%%%%%%%%%%%%%%%%%%%%%%%%%%%%%%%%%%%%%%%%%%
\begin{equation}
\begin{split}
h_{n\alpha}= & -\frac{\partial E}{\partial m_{n\alpha}}+\mu_n B_{\alpha} \\
= & \sum_{n',\beta}\left[J_{nn'}^\text{PM}\delta_{\alpha\beta}+\sum_{\gamma}D_{nn',\gamma}^\text{PM}\epsilon_{\gamma\alpha\beta}\right]m_{n'\beta} \\
& -2K\sum_{n'}\delta_{ss'}(\textbf{m}_{n}\cdot \textbf{m}_{n'})m_{n',\alpha}+\mu_n B_{\alpha},
\label{Eqh2}
\end{split}
\end{equation}
%%%%%%%%%%%%%%%%%%%%%%%%%%%%%%%%%%%%%%%%%%%%%%%%%%%%%%%%%%%%%%%%%%%%%%%%%%%%%%%%%
which directly follows from $\nabla_{\textbf{m}_n}\mathcal{G}=\textbf{0}$, i.e.\ the equilibrium condition~\cite{0305-4608-15-6-018,PhysRevB.99.144424}.
Eq.\ (\ref{Eqh2}) includes the effect of a biquadratic interaction between nearest-neighbors of the same sublattice, described by a parameter $K$ resulting from considering only a biquadratic free energy term $\frac{1}{2}K\sum_{nn'}\delta_{ss'}(\textbf{m}_{n}\cdot \textbf{m}_{n'})^2$ in Eq.\ (\ref{EQG}). Hence, and as reflected in Eq.\ (\ref{Eqh2}), positive values of $K$ favour non-collinear alignment of the order parameters.

We note that in our theory $\{\textbf{m}_n\}$ is given by a mean-field function depending on $\{\beta\textbf{h}_n\}$ only, as shown in Eq.\ (\ref{Eqm}). A solution $\{\textbf{m}_1,\dots,\textbf{m}_{N_{sc}}\}$ satisfying Eq.\ (\ref{Eqh2}), i.e.\ a magnetic state that minimizes the Gibbs free energy, for given values of $T$, $\textbf{B}$, $K$, and a particular size of the magnetic unit cell, containing $N_{sc}$ magnetic sites, can be found by using Eq.\ (\ref{Eqm}) within the following iterative method:
\begin{enumerate}
  \item We make an initial guess of $\{\textbf{m}_1^\text{in},\dots,\textbf{m}_{N_{sc}}^\text{in}\}$ that describes as close as possible the magnetic structure of interest.
  \item $\{\textbf{m}_1,\dots,\textbf{m}_{N_{sc}}\}=\{\textbf{m}_1^\text{in},\dots,\textbf{m}_{N_{sc}}^\text{in}\}$ is used as a trial input in Eq.\ (\ref{Eqh2}), from which $\{\textbf{h}_1,\dots,\textbf{h}_{N_{sc}}\}$ is calculated.
  \item The outcome of step 2, which directly gives $\{\beta\textbf{h}_n\}$ for the value of $T$ chosen, is then used to re-calculate $\{\textbf{m}_1,\dots,\textbf{m}_{N_{sc}}\}$ from Eq.\ (\ref{Eqm}).
  \item At this point $\{\textbf{m}_1^\text{in},\dots,\textbf{m}_{N_{sc}}^\text{in}\}$ and $\{\textbf{m}_1,\dots,\textbf{m}_{N_{sc}}\}$ are compared. If they are different, a new guess of $\{\textbf{m}_1^\text{in},\dots,\textbf{m}_{N_{sc}}^\text{in}\}$ is made by constructing some mixing of $\{\textbf{m}_1^\text{in},\dots,\textbf{m}_{N_{sc}}^\text{in}\}$ and $\{\textbf{m}_1,\dots,\textbf{m}_{N_{sc}}\}$.
\end{enumerate}
 Steps 2 to 4 are iteratively repeated until $\{\textbf{m}_n^\text{in}\}\approx\{\textbf{m}_n\}$ is obtained self-consistently up to a certain desired minimal error, which in our calculations was $10^{-2}$.

 The initial guesses used to obtain the self-consistent solutions for the ferromagnetic and single-q (1q) states studied in Sec.\ \ref{SecIII} were $\{\textbf{m}_\text{FM},\dots,\textbf{m}_\text{FM}\}$ and $\{\textbf{m}_n=m_\text{1q}\left(\cos(\textbf{q}_\text{1q}\cdot\textbf{R}_n)\hat{\textbf{x}}+\sin(\textbf{q}_\text{1q}\cdot\textbf{R}_n)\hat{\textbf{y}}\right)\}$, respectively. Here $m_\text{FM}$ and $m_\text{1q}$ are arbitrary values, and $\textbf{q}_\text{1q}=\frac{2\pi}{\lambda_{sc}}\hat{\textbf{z}}$, where $\lambda_{sc}$ is the the size of the supercell considered. The initial guess used to obtain the two triple-q (3q) states studied was
%%%%%%%%%%%%%%%%%%%%%%%%%%%%%%%%%%%%%%%%%%%%%%%%%%%%%%%%%%%%%%%%%%%%%%%%%%%%%%%%%%%
\begin{equation}
\begin{split}
 \label{3qguess}
    \mathbf{m}_n = 
      m_\text{3q} \Big[
     & \cos(\mathbf{q}_x\cdot\mathbf{R}_n)\,\hat{\mathbf{y}} + \sin(\mathbf{q}_x\cdot\mathbf{R}_n)\,\hat{\mathbf{z}} \\
   + & \cos(\mathbf{q}_y\cdot\mathbf{R}_n)\,\hat{\mathbf{z}} + \sin(\mathbf{q}_y\cdot\mathbf{R}_n)\,\hat{\mathbf{x}} \\
   + & \cos(\mathbf{q}_z\cdot\mathbf{R}_n)\,\hat{\mathbf{x}} + \sin(\mathbf{q}_z\cdot\mathbf{R}_n)\,\hat{\mathbf{y}} \Big],
\end{split}
\end{equation}
%%%%%%%%%%%%%%%%%%%%%%%%%%%%%%%%%%%%%%%%%%%%%%%%%%%%%%%%%%%%%%%%%%%%%%%%%%%%%%%%%%%
where $\textbf{q}_{x}=\frac{2\pi}{\lambda_{sc}}\hat{\textbf{x}}$, $\textbf{q}_{y}=\frac{2\pi}{\lambda_{sc}}\hat{\textbf{y}}$, $\textbf{q}_{z}=\frac{2\pi}{\lambda_{sc}}\hat{\textbf{z}}$, and $m_\text{3q}$ is also an arbitrary value.

\subsection{Computational DFT details}
\label{CompDetail}

All the DFT calculations were performed employing the all-electron Korringa-Kohn-Rostoker Green function (KKR-GF) package developed in Jülich~\cite{Papanikolaou2002,Bauer2014}. We used a full potential description with spin-orbit coupling added to the scalar-relativistic approximation.
Exchange and correlation effects have been treated in the local spin density approximation (LSDA) as parametrized by Vosko, Wilk and Nusair~\cite{doi:10.1139/p80-159}, the single-site scattering problem is solved with an angular momentum cutoff set to $l_{max}=3$, and the Brillouin zone is sampled with a k-mesh of $40\times 40\times 40$.
The KKR-GF method is combined with the Coherent Potential Approximation (CPA)~\cite{PhysRev.156.809,PhysRevB.5.2382,0305-4608-15-6-018} to simulate the high-temperature paramagnetic state by employing the so-called alloy analogy, with the Mn/Fe sites occupied by two magnetic species distinguished by the orientation of their spin moments (`up' and `down') and having equal concentrations, so that the average spin moment on each site vanishes.
The infinitesimal rotation method~\cite{PhysRevB.68.104436,PhysRevB.79.045209} is utilized to evaluate the pairwise magnetic interactions either in the zero-temperature ferromagnetic state ($\{J_{nn'}^\text{FM}, D_{nn',\gamma}^\text{FM}\}$) or in the high-temperature paramagnetic state ($\{J_{nn'}^\text{PM}, D_{nn',\gamma}^\text{PM}\}$).

The crystal structure of MnGe belongs to the space group of $P2_13$ (198) where both Mn and Ge atoms sit in 4a Wyckoff positions~\cite{Dyadkin}.
In our calculations we have used $\textbf{r}_\text{Mn}=(0.8627,0.8627,0.8627)$ and $\textbf{r}_\text{Ge}=(0.15652,0.15652,0.15652)$, respectively, which generates a primitive cell containing four Mn atoms and four Ge atoms. For simplicity and comparison the same internal coordinates were used for FeGe.

\section{Temperature-dependent magnetism of B20 materials}
\label{SecIII}
\subsection{Non-canonical mechanism for the short period of MnGe}

%%%%%%%%%%%%%%%%%%%%%%%FIGURE%%%%%%%%%%%%%%%%%%%%%%%%%%%%%%%%%%%%%
\begin{figure}[tb]
\centering
\includegraphics[width=\columnwidth]{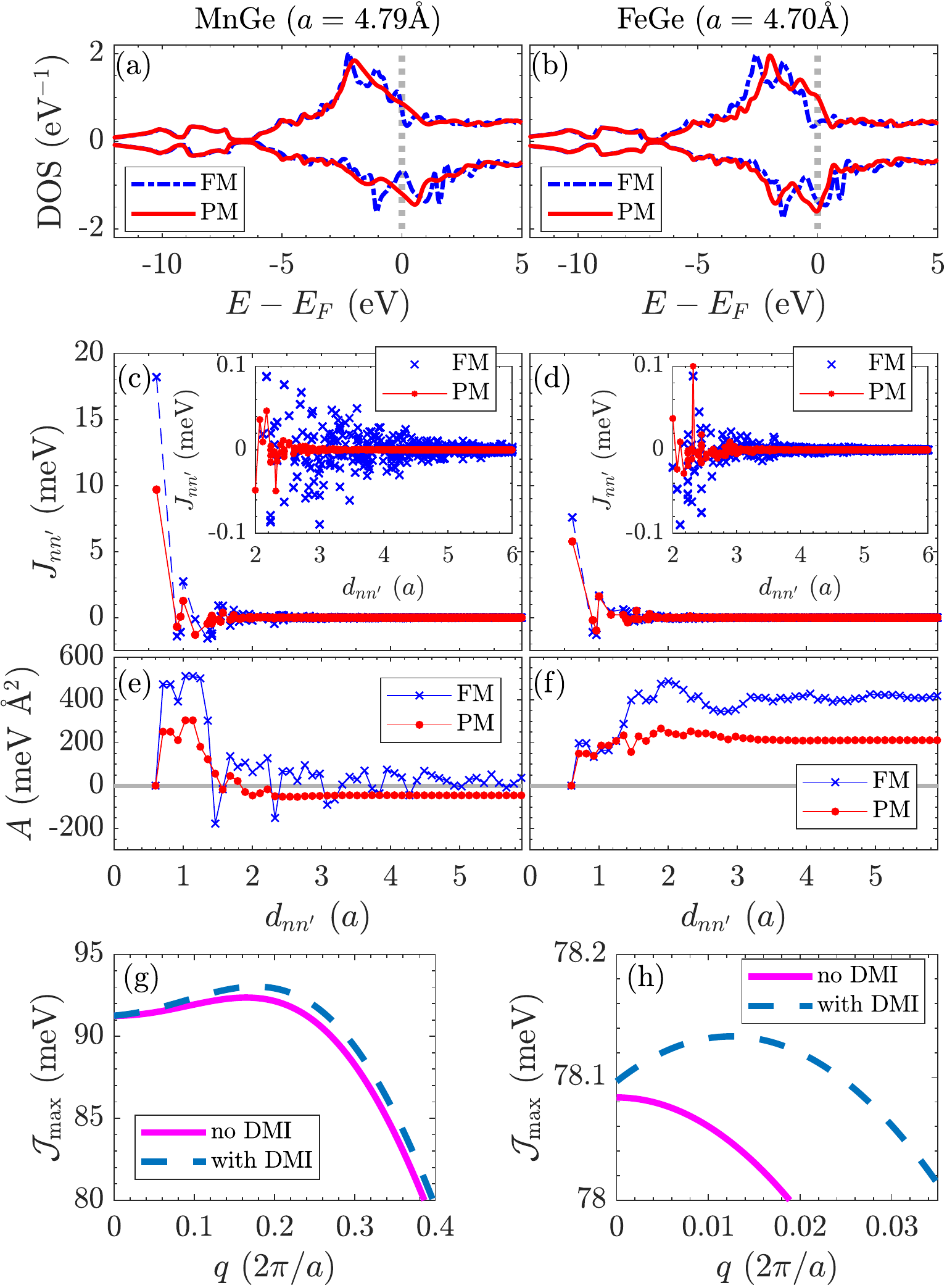}
\caption{
Electronic structure and magnetic interactions of MnGe (left panels, (a,c,e,g)) and FeGe (right panels, (b,d,f,h)) in the ferromagnetic and paramagnetic states.
The paramagnetic state is obtained by employing the coherent potential approximation, see Sec.\ \ref{CompDetail}.
(a,b) Density of states for an electron spin-polarized parallel (positive) and anti-parallel (negative) to the direction of the local moment on the site.
(c,d) Pairwise isotropic interactions, %obtained for MnGe ($a=4.79$\AA) and FeGe ($a=4.70$\AA) in their FM and PM phases 
obtained in the ferromagnetic ($J_{nn'}^\text{FM}$) and paramagnetic ($J_{nn'}^\text{PM}$) states,
as functions of the distance between the two atoms.
The insets show a magnified view of the results for larger distances. 
(e,f) Dependence of the spin stiffness $A = \sum_{n'} J_{nn'} d_{nn'}^2$ on the cutoff distance for the sum.
(g,h) Solid lines: The largest eigenvalue of the interaction matrix obtained by lattice Fourier transforming $J_{nn'}^\mathrm{PM}$, i.e.\ $J^\text{PM}(\textbf{q})$,
as function of the modulus of the wave vector along the [111] direction.
A peak in this quantity signifies the dominant magnetic instability of the paramagnetic state, in connection to Eq.~\eqref{eq:free_high_temp}.
Dashed lines: Including the DMI in the interaction matrix, i.e.\ largest eigenvalue of $\mathcal{J}(\textbf{q})$.
See Sec.\ \ref{Fourier} for further details.
}%
\label{Fig1}
\end{figure}
%%%%%%%%%%%%%%%%%%%%%%%%%%%%%%%%%%%%%%%%%%%%%%%%%%%%%%%%%%%%%%%%%%%%

We first study the electronic structures of MnGe and FeGe constrained to both the ferromagnetic and paramagnetic states.
Fig.\ \ref{Fig1}(a,b) compares the density of states (DOS) obtained in these two limits, which we have found to be fairly similar in both MnGe and FeGe.
Magnetic disorder causes an effective broadening of the electronic bands, which leads to a smearing of the DOS in the paramagnetic state, and a redistribution of spectral weight in comparison to the ferromagnetic state.
Due to the latter, the local spin moments are smaller in the paramagnetic state than in the ferromagnetic state:  $1.8\,\mu_\mathrm{B}$ vs.\ $2.1\,\mu_\mathrm{B}$ for MnGe and $0.8\,\mu_\mathrm{B}$ vs.\ $1.2\,\mu_\mathrm{B}$ for FeGe, respectively.
Fig.\ \ref{Fig1}(c,d) shows the dependence of the $J_{nn'}$ on the distance $d_{nn'}$ between the atoms (in units of the lattice constant $a$) for the ferromagnetic ($J_{nn'}^\text{FM}$) and paramagnetic ($J_{nn'}^\text{PM}$) states, while panels (e,f) show the corresponding spin stiffness $A = \sum_{n'} J_{nn'} d_{nn'}^2$ as a function of the cutoff distance for the sum.
As is common in itinerant magnets, the interactions are long-ranged and oscillate with distance. Their long-distance behavior reveals the impact of the state of magnetic order on the electronic structure.
The magnetic disorder characterizing paramagnetism hinders the propagation of the conduction electrons that mediate the long-range interactions, leading to a much faster decay with distance than found for the ferromagnetic state.
This is especially clear for the spin stiffness, which converges quickly in the paramagnetic state and quite poorly in the ferromagnetic state.
We also see that the asymptotic limit of $A$ is different for each state: For FeGe there is a factor of two between the two limiting values of $A$, while for MnGe the paramagnetic state gives $A < 0$ while the ferromagnetic state gives $A \sim 0$.
Overall, this shows that the magnetic interactions derived from the paramagnetic state can lead to very different predictions than those derived from the ferromagnetic state.

At sufficiently high temperature any magnetic phase disorders and transforms into the paramagnetic state.
Hence, knowledge of a single set of high-temperature interactions, $\{J_{nn'}^\mathrm{PM},\mathbf{D}_{nn'}^\mathrm{PM}\}$, enables to generally predict the potential instabilities of the paramagnetic state to the formation of any type of magnetic ordering for a given material~\cite{0305-4608-15-6-018,PhysRevLett.93.257204,PhysRevB.99.144424}.
$\{J_{nn'}^\mathrm{FM},\mathbf{D}_{nn'}^\mathrm{FM}\} \neq \{J_{nn'}^\mathrm{PM},\mathbf{D}_{nn'}^\mathrm{PM}\}$, which means that in general this role cannot be played by interactions computed from the ferromagnetic state.
Moreover, to compare the relative stability of zero-temperature magnetic states one needs to calculate the total energies of all of them.

%%%%%%%%%%%%%%%%%%%%%%%FIGURE%%%%%%%%%%%%%%%%%%%%%%%%%%%%%%%%%%%%%
\begin{figure}
\centering
\includegraphics[clip,scale=0.69]{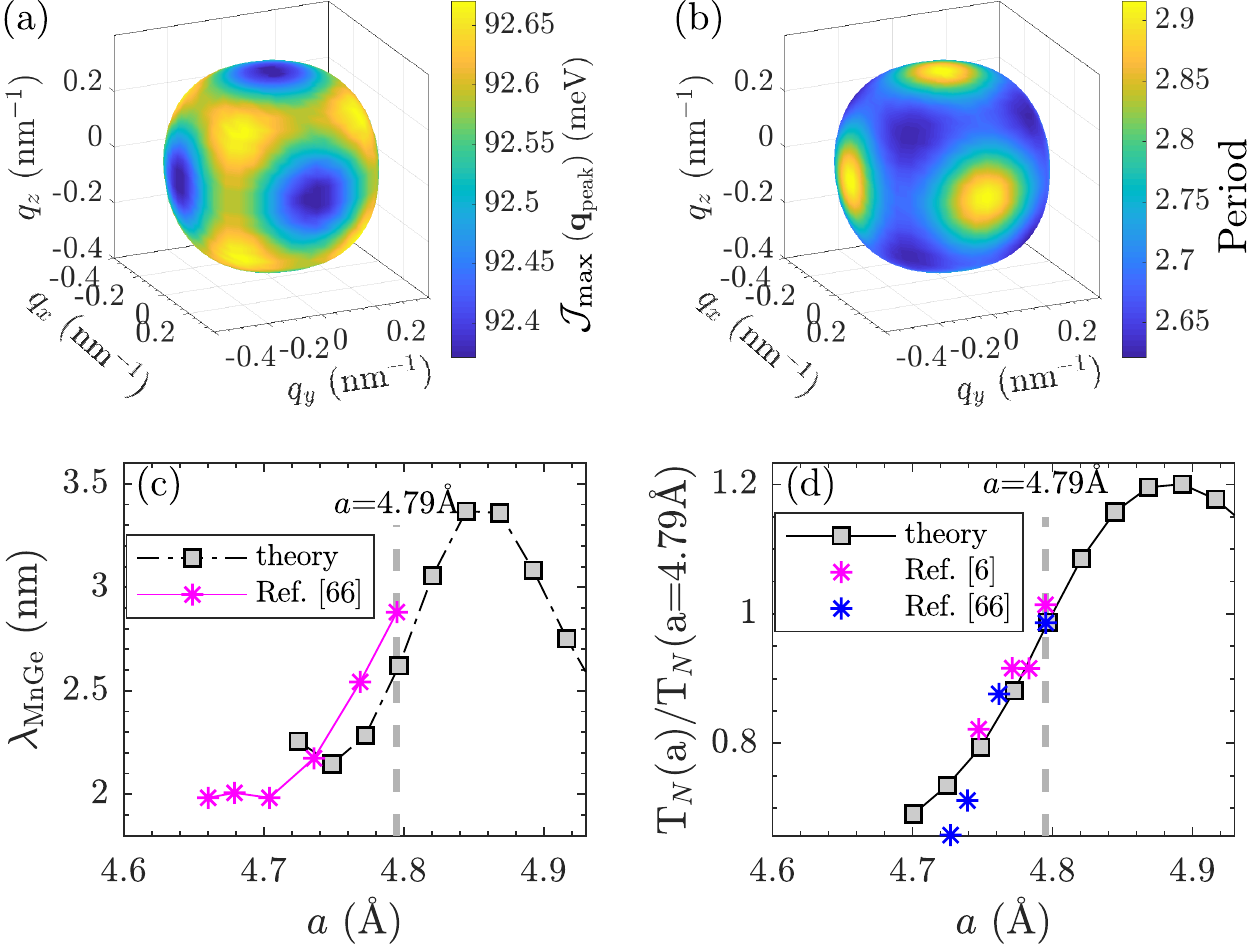}
\caption{
Helical instabilities of the paramagnetic state of MnGe from isotropic paramagnetic interactions $J_{nn'}^\mathrm{PM}$.
(a,b) Surface plots in reciprocal space of $\mathcal{J}_\mathrm{max}(\mathbf{q}_\mathrm{peak})$ (without DMI) and of $\lambda_\mathrm{MnGe}(\mathbf{q}_\mathrm{peak})$, respectively, with the distance to the centre being the magnitude $q_\mathrm{peak}$ of the wave vector at which the peak of $\mathcal{J}_\mathrm{max}(\mathbf{q}_\mathrm{peak})$ is found for $a=\SI{4.79}{\angstrom}$.
(c) Magnetic period and relative N\'eel transition temperature obtained for the most stable magnetic phase at different values of the lattice parameter, $a$.
Results are compared with available experimental data~\cite{MnGeNat,PhysRevB.89.180407}, which has been extracted by considering how $T_\mathrm{N}$, $\lambda_\mathrm{MnGe}$, and $a$ change against pressure.
}
\label{Fig2}
\end{figure}
%%%%%%%%%%%%%%%%%%%%%%%%%%%%%%%%%%%%%%%%%%%%%%%%%%%%%%%%%%%%%%%%%%%%

In the paramagnetic limit the magnetic free energy in Eq.~\eqref{eq:free_high_temp} is a quadratic form of the order parameters. We minimize $\mathcal{G}$ in this limit by diagonalizing the lattice Fourier transform of the interactions, as explained in Sec.\ \ref{Fourier}. The largest eigenvalue of the total interaction matrix (including both $J_{nn'}^\mathrm{PM}$ and $\mathbf{D}_{nn'}^\mathrm{PM}$), $\mathcal{J}_\mathrm{max}(\mathbf{q})$, indicates the dominant magnetic instability of the paramagnetic state.
Fig.\ \ref{Fig1}(g,h) shows $\mathcal{J}_\mathrm{max}(\mathbf{q})$ along the [111] direction, with and without the DMI.
The magnetic instability is characterized by a wave vector $\mathbf{q}_\mathrm{peak}$ indicating the peak location in $\mathcal{J}_\mathrm{max}(\mathbf{q})$ and the spatial period of the magnetic modulation $\lambda = 2\pi/q_\mathrm{peak}$.
FeGe follows the canonical picture of B20 helimagnetism~\cite{Bak1980}: without DMI the ferromagnetic state is the dominant instability ($q_\mathrm{peak} = 0$), with the DMI stabilizing a long-period helimagnetic structure with $\lambda_\mathrm{FeGe} = \SI{40}{\nano\meter}$ (the experimental value is \SI{70}{\nano\meter}~\cite{Lebech_1989}).
The behavior of MnGe is strikingly different: without DMI the long-ranged isotropic interactions already cooperate to stabilize instead a helical state with a very short period, with the DMI making a minor correction only.
The calculated period $\lambda_\text{MnGe}=\SI{2.6}{\nano\meter}$ is in good agreement with the experimental range $3-\SI{6}{\nano\meter}$~\cite{MnGeNat}.
Our high-temperature approach to the magnetic instabilities of FeGe and MnGe offers a consistent physical picture, even though the dominant magnetic interactions are very different for each material.
In sharp contrast, using the interactions computed from the ferromagnetic state we find a helical period of \SI{196}{\nano\meter} for FeGe, consistent with previous results~\cite{PhysRevB.100.214406}, and MnGe to have a ferromagnetic instability which the DMI turns into a helix of period \SI{40}{\nano\meter}.

We now center our attention on the magnetic properties of MnGe as given by the dominant isotropic interactions obtained in the paramagnetic state $J_{nn'}^\mathrm{PM}$, i.e.\ $\mathcal{J}_\mathrm{max}(\mathbf{q})$ without the DMI.
It turns out that $\mathcal{J}_\mathrm{max}(\mathbf{q})$ peaks along all possible directions in reciprocal space with similar values of $q_\mathrm{peak}$ and very similar energies, i.e.\ the corresponding helical instabilities are nearly degenerate.
This is illustrated in Fig.\ \ref{Fig2}(a), where we plot a surface whose distance to the centre is $q_\mathrm{peak}$ along the corresponding direction and its color encodes $\mathcal{J}_\mathrm{max}(\mathbf{q}_\text{peak})$.
Similarly, Fig.\ \ref{Fig2}(b) shows $\lambda_\mathrm{MnGe}$, and from its range of variation we see that this surface is almost spherical.
%This figure shows that indeed there is a very little change of $\mathcal{FT}[J_{nn'}^\mathrm{PM}](\textbf{q})$ for different values of $\mathbf{q}_\mathrm{peak}$.
We have systematically found that the highest peak is along the \textless111\textgreater\ directions. However, the very small energy cost of changing to a different direction means that other instabilities could be favored by other type of interactions, such as the magnetocrystalline anisotropy (not considered here), and higher than pairwise ones, as will be shown later.

%%%%%%%%%%%%%%%%%%%%%%%FIGURE%%%%%%%%%%%%%%%%%%%%%%%%%%%%%%%%%%%%%
\begin{figure*}
\centering
\includegraphics[clip,scale=0.68]{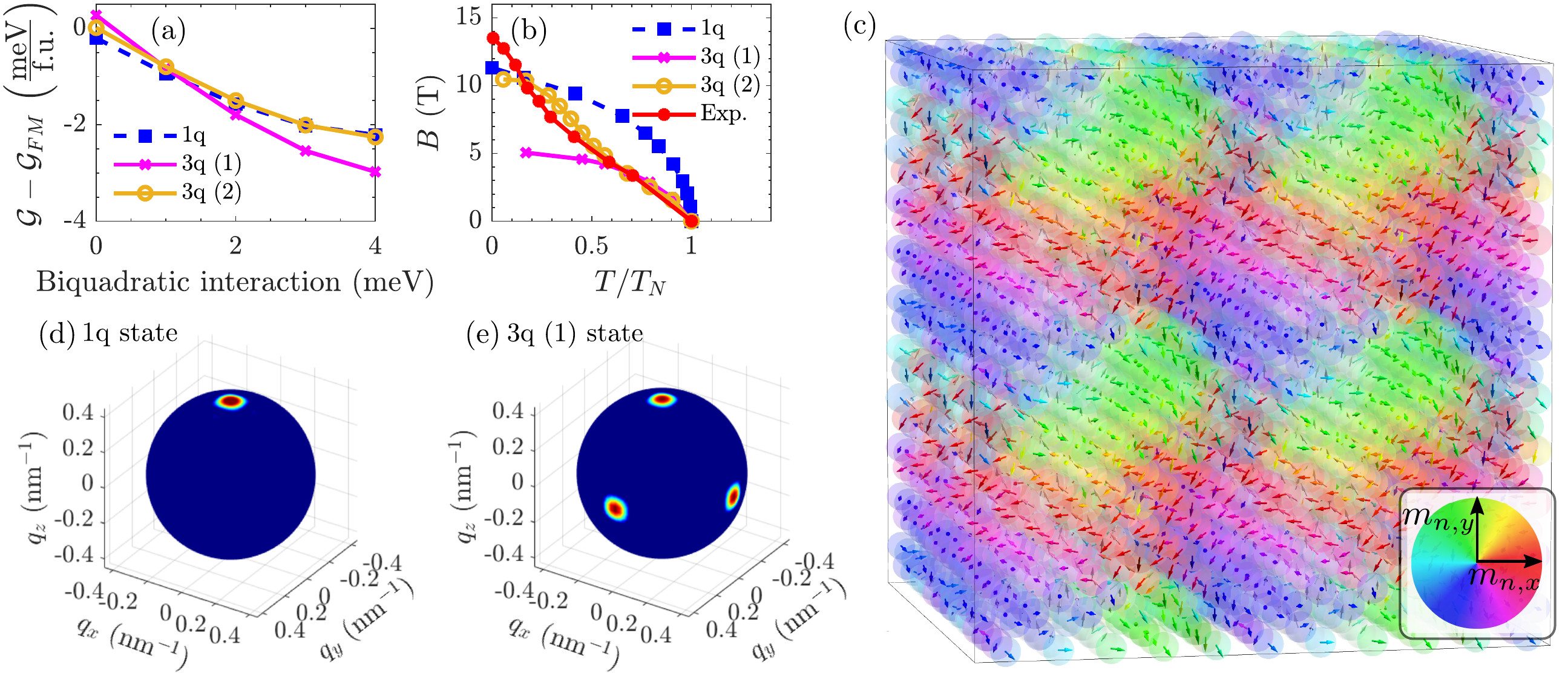}
\caption{
Free energy of MnGe including a biquadratic interaction, $K$, for the stabilization of 3q-states.
(a) Dependence of the free energies of 1q and 3q states, relative to the ferromagnetic one ($\mathcal{G}_{FM}$), on the added biquadratic interaction, at $T=20$K and $B=0$.
(b) Critical magnetic field needed to stabilize a ferromagnetic state, both from solely isotropic pairwise interactions ($K=0$), which stabilizes the 1q state, and when a biquadratic interaction of strength $K=4$ meV is added, which stabilizes at least two different 3q-states in an applied magnetic field.
Experimental data shown is extracted from Ref.\ ~\cite{PhysRevB.88.064409}.
(c) Representation of the real-space averaged orientations, $\{\textbf{m}_n\}$, obtained for the 3q (1) state, the most stable one when the biquadratic interaction $K=4$ meV is added at $T=20$K and $B=0$.
The $x$ and $y$ components are given by the color wheel, while the out-of-plane component shades the arrows towards black.
(d) Spherical cut of the intensity of the corresponding total lattice Fourier transform of the helimagnetic 1q and 3q (1) states, $m_\text{1q}(\textbf{q}_{cut})$ and $m_\text{3q(1)}(\textbf{q}_{cut})$, respectively (see text). The more red the color is, the larger the intensity is. 
}%
\label{Fig3}
\end{figure*}
%%%%%%%%%%%%%%%%%%%%%%%%%%%%%%%%%%%%%%%%%%%%%%%%%%%%%%%%%%%%%%%%%%%%

The minimization of Eq.~\eqref{eq:free_high_temp} provides the mean-field estimate $3k_\mathrm{B}T_\mathrm{N} = \mathcal{J}_\mathrm{max}(\mathbf{q}_\mathrm{peak})$ through Eq.\ (\ref{Eq3}).
This gives $T_\mathrm{N}= \SI{360}{\kelvin}$ at $a=\SI{4.79}{\angstrom}$, substantially above the experimental value of \SI{180}{\kelvin}~\cite{MnGeNat}.
This large overestimation of $T_\mathrm{N}$ in the mean-field approximation points to the importance of non-local spin fluctuations in MnGe.
We have carried out Monte Carlo simulations using the set of magnetic interactions obtained from the paramagnetic state~\footnote{We use an in-house Monte Carlo code that considers isotropic exchange interactions and the Metropolis algorithm~\cite{PhysRevB.88.134403,10.1145/272991.272995}. In our simulations we considered a $18\times 18\times 18$ supercell (i.e.\ 23328 magnetic atoms) and for each temperature the number of Monte Carlo steps was $10^{6}$. From this we obtained $10^{5}$ sampling events equally spaced by 10 steps after a relaxation time of $10^{4}$ steps. The determination of the transition temperature was done through the calculation and examination of the heat capacity.}, which indicate an estimate close to $T_\mathrm{N}^\text{MC} = \SI{160}{\kelvin}\,(J_{nn'}^\text{PM})$ in good agreement with experiment.
This is in stark contrast to previous simulations based on the ferromagnetic state, which led to a value of $\SI{300}{\kelvin}\,(J_{nn'}^\text{FM})$~\cite{Bornemann_2019} that strongly overestimates the experimental one.
It seems that in order to find a good estimate of $T_\mathrm{N}$ for MnGe both non-local spin fluctuations and the set of magnetic interactions obtained in the appropriate high-temperature paramagnetic limit must be used.
 Work  is  inprogress to fully corroborate this hypothesis. 
Another factor that can impact the value of $T_\mathrm{N}$ is its strong dependence on the lattice parameter, which we now address.

The results discussed so far were computed for the experimental value $a=\SI{4.79}{\angstrom}$, but for MnGe the magnetic properties such as the spin moment, $\lambda_\mathrm{MnGe}$, and $T_\mathrm{N}$ depend strongly on the lattice constant~\cite{MnGeNat,PhysRevB.89.180407,PhysRevB.93.214404}.
Fig.\ \ref{Fig2}(c,d) shows that our results reproduce this strong dependence.
In order to compare the trends with experimental data in panel (d), we plot $T_\mathrm{N}(a)$ normalized to the value obtained at $a=\SI{4.79}{\angstrom}$.
We observe a linear dependence of $T_\mathrm{N}(a)$ for $a$ between \SI{4.75}{\angstrom} and \SI{4.85}{\angstrom}, of approximately a 30\% decrease of $T_\mathrm{N}$ by a \SI{0.07}{\angstrom} reduction of $a$.
The horizontal offset between theory and experiment is likely connected to inaccuracies in determining the lattice constant within DFT.
The experimental comparison made in Fig.\ \ref{Fig2}(c,d)~\cite{MnGeNat,PhysRevB.89.180407} pleasingly shows that our theory correctly captures trends qualitatively and sometimes quantitatively.

\subsection{Stability of a triple-q magnetic state}

The pairwise interactions obtained in the high-temperature paramagnetic limit ($\{J_{nn'}^\mathrm{PM},\textbf{D}_{nn'}^\mathrm{PM}\}$) govern the energetics and behavior of $T_\mathrm{N}$ and the period of single-q and multiple-q magnetic states~\cite{Chaikin}, which explains the good agreement between our results and experimental data shown in Figs.\ \ref{Fig1} and \ref{Fig2}.
However, experiments have also shown that the ground-state magnetic structure is not a single-q (1q) helix but a complex triple-q (3q) state~\cite{PhysRevLett.106.156603,MnGeNat}, which demands the inclusion of higher-order magnetic interactions in the free energy~\cite{Chaikin}.
While the $J_{nn'}^\mathrm{PM}$ select the most energetically favorable wave vectors ($\{\mathbf{q}_\mathrm{peak}\}$), higher-order magnetic interactions can couple several wave vectors from this set and stabilize a multiple-q state~\cite{Binz2006a,Okumura2020}.
The present computational approach can be used to study the potential effect of such higher-order interactions
on the stabilization of multiple-q states. In the following we explore how they might change the magnetism of MnGe by adding a biquadratic interaction between nearest-neighbors of the same sublattice, $K$, to the scenario of isotropic pairwise interactions $J_{nn'}^\mathrm{PM}$.
To study this effect we minimize the Gibbs free energy of MnGe at different values of $K$, as described in Sec.\ \ref{min}.

Fig.\ \ref{Fig3}(a) shows the free energies of 1q helimagnetic and two different 3q states (3q (1) and 3q (2)) at $T=\SI{20}{\kelvin}$, relative to the free energy of the ferromagnetic state.
These states are very close in energy when considering solely $J_{nn'}^\mathrm{PM}$, so a small value of the additional biquadratic interaction is enough to make a 3q state the most stable one.
The 3q states are obtained by iterative minimization of the free energy, starting from a superposition of three 1q states with wave vectors along each of the cubic directions (see Eq.\ \ref{3qguess}), as explained in Sec.\ \ref{min}.
To this end, we used a $5\times 5\times 5$ supercell, which can describe 1q and 3q states of periods, $\lambda_{sc}$, very close to the ones predicted in Fig.\ \ref{Fig1}. 
We point out that the 3q (1) state converged when the initial guess described in Eq.\ (\ref{3qguess}) was used for low values of the magnetic field. For larger magnetic field values, the 3q (2) state converged instead, whose solution was used as an initial iterative guess to obtain a self-consistent solution at zero magnetic field.

The complex magnetic structure $\{\textbf{m}_n\}$ obtained for the most stable 3q (1) state at zero external magnetic field is shown in Fig.~\ref{Fig3}(c). We confirmed its 3q nature by calculating its total lattice Fourier transform, $m_\text{3q(1)}(\textbf{q})$, which is obtained by applying Eq.\ (\ref{mqb}) to the magnetic order parameters obtained for this state at $T=20$K and $B=0$. A spherical cut of $m_\text{3q(1)}(\textbf{q}_{cut})$  for values of the wave vector describing periods of the size of the supercell, i.e.\ $q_{cut}=\frac{2\pi}{\lambda_{sc}}$, is shown in Fig.~\ref{Fig3}(e), which demonstrates that the corresponding Fourier transform shows three singularities along the three crystallographic directions. Moreover, the Fourier transform of the 1q-state, $m_\text{1q}(\textbf{q}_{cut})$ (Fig.~\ref{Fig3}(d)), shows its different single-q nature.

The temperature dependence of the critical magnetic field needed to stabilize the ferromagnetic state provides the last stringent test of our results.
Considering only isotropic pairwise interactions leads to a zero-temperature value of the critical field which is already in very good agreement with experiment~\cite{PhysRevB.88.064409,MnGeNat}, as seen for the 1q-to-ferromagnetic transition in Fig.\ \ref{Fig3}(b) (dashed line) obtained for $K=0$ meV and an applied magnetic field perpendicular to the basal plane of the helimagnetic state~\footnote{We point out that the iterative method to minimize the Gibbs free energy explained in Sec.\ \ref{min} was used to generate the data shown in Fig.\ \ref{Fig3}(b) for the  3q (1) and 3q (2) states, at $K=4$ meV. The calculation of the critical line corresponding to the 1q state (at $K=0$ meV) is made by minimizing instead a simplified form of the Gibbs free energy describing an arbitrary value of the wave vector $\textbf{q}$, which is explained in appendix \ref{1q}. Such an expression of $\mathcal{G}$, shown in Eq.\ \ref{Gfouu}, is obtainable for single-q states only.}.
However, adding a biquadratic interaction transforms the low-field helimagnetic state into a 3q-state and also can change the upper critical field line to have a shape that resembles more closely the experimental one, while keeping a fair quantitative agreement.
This is shown by the 3q (1) and 3q (2) critical lines in Fig.\ \ref{Fig3}(b) obtained for a value of $K=4$ meV and an applied magnetic field along the [111] direction.
Our calculations also suggest that intermediate magnetic phases might be present, as indicated by the lower critical field line and the potential stabilization of more than one 3q-state.
The critical lines for the two 3q (1) and 3q (2) states were obtained by finding the values of $T$ and $B$ at which the iterative minimization of the Gibbs free energy does not converge. This means that these describe first-order (discontinuous) magnetic phase transitions from 3q states to the FM state.
We highlight that these results are meant to support the quality of our pairwise magnetic interactions and not to present the definite picture of the magnetic phase diagram of MnGe.
A more realistic set of higher-order interactions could lead to a richer diagram and to closer agreement with experimental findings.

\section{Conclusions}
\label{SecConc}

To summarize, we used a finite-temperature first-principles theory to demonstrate the dominant role of isotropic magnetic interactions for the short-period magnetism of MnGe, in stark contrast with the canonical picture of B20 helimagnetism. Our approach exploits the fact that pairwise interactions obtained in the high-temperature paramagnetic phase of a given material can be used to generally describe its instabilities to the formation of wave-modulated magnetic states, and their periods and transition temperatures as functions of the lattice structure.
We applied the theory to both MnGe and FeGe (a canonical helimagnet) and obtained good agreement with various experimental results in both cases.
For MnGe, the calculation and minimization of its free energy has showed that rather weak higher-order interactions are enough to stabilize a 3q state instead of a helical state at lower temperatures
and have a significant influence on the temperature dependence of the critical magnetic field that stabilizes the ferromagnetic state.
The details of the nature and strength of the higher-order magnetic interactions for MnGe are beyond the scope of our work, but recent developments show promise in this respect~\cite{PhysRevB.99.144424,Brinker2019,Mankovsky2020,brinker2020prospecting,SergiiNatComm,Lounis2020}.
%We highlight that recent developments have pointed out ways to access higher-order magnetic interactions from first-principles~\cite{PhysRevB.99.144424,Brinker2019,Mankovsky2020,brinker2020prospecting}, which could supply the missing microscopic information on the nature and strength of such interactions to fully describe MnGe at intermediate and low temperatures~\cite{SergiiNatComm}.

\begin{acknowledgements}
The authors gratefully acknowledge fruitful discussions with N.\ Kanazawa, B.\ Zimmermann and S.\ Brinker as wellas P. Mavropoulos for sharing his Monte Carlo code with us.
E.\ Mendive-Tapia acknowledges funding from the DAAD Short-Term Grants (2019) and the priority programme SPP1599 ``Ferroic Cooling'' (Grant No.\ HI1300/6-2).
E.\ Mendive-Tapia, M.\ dos Santos Dias, S.\ Lounis, and S.\ Bl\"ugel are grateful for the support from the European Research Council (ERC) under the European Union's Horizon 2020 research and innovation programme (ERC-consolidator Grant No.\ 681405 - DYNASORE, and ERC-synergy grant No.\ 856538 - 3D MAGiC).
We acknowledge financial supportfrom  the  DARPA  TEE  program  through  grant  MIPR  (No.\ HR0011831554)  from  DOI,  from  Deutsche  Forschungsgemeinschaft (DFG) through SPP 2137 ``Skyrmionics'' (Project BL  444/16),  the  Collaborative  Research  Centers  SFB  1238 (Project  C01)  as  well  as  computing  resources  at  the  supercomputers  JURECA  at  Juelich  Supercomputing  Centre and  JARA-HPC  from  RWTH  Aachen  University  (Projects jara0224  and  jara3dmagic).
J.B.S.  acknowledges  fundingfrom EPSRC (UK) Grant No. EP/M028941/1.
\end{acknowledgements}

\appendix

\section{Helimagnetic state and its spatial dependence}
\label{Modulation}

The single-site internal magnetic fields in Eq.\ (\ref{EqH0}) are effective fields sustaining the magnetic moments for a given amount of magnetic order $\{\textbf{m}_n\}$ and are obtained by computing the first derivative of the internal magnetic energy with respect to the local order parameter~\cite{0305-4608-15-6-018}. In the paramagnetic limit ($\{\textbf{m}_n\rightarrow\textbf{0}\}$) and setting $\textbf{B}=\textbf{0}$, Eq.\ (\ref{Eqh2}) becomes
%%%%%%%%%%%%%%%%%%%%%%%%%%%%%%%%%%%%%%%%%%%%%%%%%%%%%%%%%%%%%%%%%%%%%%%%%%%%%%%%%%%
\begin{equation}
h_{n\alpha}=-\frac{\partial E}{\partial m_{n\alpha}}=\sum_{n',\beta}\left[J_{nn'}^\text{PM}\delta_{\alpha\beta}+\sum_{\gamma}D_{nn',\gamma}^\text{PM}\epsilon_{\gamma\alpha\beta}\right]m_{n'\beta}.
\label{Eqh}
\end{equation}
%%%%%%%%%%%%%%%%%%%%%%%%%%%%%%%%%%%%%%%%%%%%%%%%%%%%%%%%%%%%%%%%%%%%%%%%%%%%%%%%%ç
It can be shown from Eq.\ (\ref{Eqm}) that 
%%%%%%%%%%%%%%%%%%%%%%%%%%%%%%%%%%%%%%%%%%%%%%%%%%%%%%%%%%%%%%%%%%%%%%%%%%%%%%%%%%%
\begin{equation}
\textbf{m}_n\approx \frac{\beta}{3}\textbf{h}_n
\label{mPM}
\end{equation}
%%%%%%%%%%%%%%%%%%%%%%%%%%%%%%%%%%%%%%%%%%%%%%%%%%%%%%%%%%%%%%%%%%%%%%%%%%%%%%%%%
for $\textbf{h}_n\rightarrow\textbf{0}$ (i.e.\ for $\textbf{m}_n\rightarrow\textbf{0}$). Using Eq.\ (\ref{Eqh}) in this paramagnetic limit, Eq.\ (\ref{mPM}) can be written as
%%%%%%%%%%%%%%%%%%%%%%%%%%%%%%%%%%%%%%%%%%%%%%%%%%%%%%%%%%%%%%%%%%%%%%%%%%%%%%%%%%%
\begin{equation}
3k_\mathrm{B} T m_{n\alpha}=\sum_{n',\beta}\left[J_{nn'}^\text{PM}\delta_{\alpha\beta}+\sum_{\gamma}D_{nn',\gamma}^\text{PM}\epsilon_{\gamma\alpha\beta}\right]m_{n'\beta},
\label{EqComp0}
\end{equation}
%%%%%%%%%%%%%%%%%%%%%%%%%%%%%%%%%%%%%%%%%%%%%%%%%%%%%%%%%%%%%%%%%%%%%%%%%%%%%%%%%
which shows that the components of the eigenvectors of the matrix form of $\mathcal{J}_{nn'\alpha\beta}=J_{nn'}^\text{PM}\delta_{\alpha\beta}+\sum_{\gamma}D_{nn',\gamma}^\text{PM}\epsilon_{\gamma\alpha\beta}$ (Eq.\ (\ref{TotalMatrix})) directly provide the spatially-dependent components of the magnetic order parameters. For its largest eigenvalue $3k_\text{B}T_\text{max}$, corresponding to the largest transition temperature, they describe how the helimagnetic phase stabilizing just below $T_\text{max}$ is modulated.
Similarly, the eigenvector components obtained from the diagonalization presented in Eq.\ (\ref{EqM}) in the Fourier space, $\{V_{s\alpha,p}(\textbf{q})\}$, contain information to express the spatial dependence of $\{m_{ts\alpha}\}$. Applying the Fourier transform to Eq.\ (\ref{EqComp0}) gives
%%%%%%%%%%%%%%%%%%%%%%%%%%%%%%%%%%%%%%%%%%%%%%%%%%%%%%%%%%%%%%%%%%%%%%%%%%%%%%%%%%%
\begin{equation}
\begin{split}
 & 3k_\mathrm{B} T m_{s\alpha}(\textbf{q})= \\ 
 & \sum_{s',\beta}\left[J_{ss'}^\text{PM}(\textbf{q})\delta_{\alpha\beta}+\sum_{\gamma}D_{ss',\gamma}^\text{PM}(\textbf{q})\epsilon_{\gamma\alpha\beta}\right]m_{s'\beta}(\textbf{q}),
\label{EqComp0q}
\end{split}
\end{equation}
%%%%%%%%%%%%%%%%%%%%%%%%%%%%%%%%%%%%%%%%%%%%%%%%%%%%%%%%%%%%%%%%%%%%%%%%%%%%%%%%%
where
%%%%%%%%%%%%%%%%%%%%%%%%%%%%%%%%%%%%%%%%%%%%%%%%%%%%%%%%%%%%%%%%%%%%%%%%%%%%%%%%%%%
\begin{equation}
m_{s\alpha}(\textbf{q})=\frac{1}{N_c}\sum_t m_{st\alpha}\exp[-i\textbf{q}\cdot\textbf{R}_t]
\label{mq}
\end{equation}
%%%%%%%%%%%%%%%%%%%%%%%%%%%%%%%%%%%%%%%%%%%%%%%%%%%%%%%%%%%%%%%%%%%%%%%%%%%%%%%%%
is the Fourier transform of the order parameter. We define the total Fourier transform of the order parameter as
%%%%%%%%%%%%%%%%%%%%%%%%%%%%%%%%%%%%%%%%%%%%%%%%%%%%%%%%%%%%%%%%%%%%%%%%%%%%%%%%%%%
\begin{equation}
m(\textbf{q})=\left|\sum_{s\alpha}m_{s\alpha}(\textbf{q})\right|.
\label{mqb}
\end{equation}
%%%%%%%%%%%%%%%%%%%%%%%%%%%%%%%%%%%%%%%%%%%%%%%%%%%%%%%%%%%%%%%%%%%%%%%%%%%%%%%%%

Eqs.\ (\ref{EqM}) and (\ref{EqComp0q}) show that the components of the eigenvectors, $\{V_{s\alpha,p}(\textbf{q})\}$, can be used to describe the wave modulation of $\{\textbf{m}_n\}$ through Eq.\ (\ref{mq}) for a given solution of the eigenvalue problem.
We first point out that $\{V_{s\alpha,p}(\textbf{q})\}$ can be complex numbers as a result of the lattice Fourier transform,
%%%%%%%%%%%%%%%%%%%%%%%%%%%%%%%%%%%%%%%%%%%%%%%%%%%%%%%%%%%%%%%%%%%%%%%%%%%%%%%%%%%
\begin{equation}
m_{s\alpha,p}(\textbf{q})=m_{0\alpha}V_{s\alpha,p}(\textbf{q})=m_{0\alpha}(\Re[V_{s\alpha,p}]+i\Im[V_{s\alpha,p}]),
\label{Vim}
\end{equation}
%%%%%%%%%%%%%%%%%%%%%%%%%%%%%%%%%%%%%%%%%%%%%%%%%%%%%%%%%%%%%%%%%%%%%%%%%%%%%%%%%%%
where $\Re[V_{s\alpha,p}]$ and $\Im[V_{s\alpha,p}]$ are the real and imaginary parts of the eigenvector, respectively, and a proportional factor, $m_{0,\alpha}$, must be added to describe different sizes of the order parameter since the eigenvector is normalized.
To show how Eq.\ (\ref{Vim}) can be interpreted we write a general form of a single-q helimagnetic structure in a multi-sublattice system,
%%%%%%%%%%%%%%%%%%%%%%%%%%%%%%%%%%%%%%%%%%%%%%%%%%%%%%%%%%%%%%%%%%%%%%%%%%%%%%%%%%%
\begin{equation}
\left\{m_{ts\alpha}=A_{s\alpha,p}\cos(\textbf{q}\cdot\textbf{R}_t)+B_{s\alpha,p}\sin(\textbf{q}\cdot\textbf{R}_t)\right\},
\label{EqGen}
\end{equation}
%%%%%%%%%%%%%%%%%%%%%%%%%%%%%%%%%%%%%%%%%%%%%%%%%%%%%%%%%%%%%%%%%%%%%%%%%%%%%%%%%%%
where the second term in the right hand side accounts for possible sublattice-dependent phase shifts. Substituting Eq.\ (\ref{EqGen}) into Eq.\ (\ref{mq}), and using the trigonometric properties
%%%%%%%%%%%%%%%%%%%%%%%%%%%%%%%%%%%%%%%%%%%%%%%%%%%%%%%%%%%%%%%%%%%%%%%%%%%%%%%%%%%
\begin{equation}
\begin{split}
 & \sum\limits_t \cos^2(\textbf{q}\cdot\textbf{R}_t)=\sum\limits_t \sin^2(\textbf{q}\cdot\textbf{R}_t)=\frac{1}{2}N_c \\
 & \sum\limits_t \cos(\textbf{q}\cdot\textbf{R}_t)\sin(\textbf{q}\cdot\textbf{R}_{t})=0
\label{util1}
\end{split}
\end{equation}
%%%%%%%%%%%%%%%%%%%%%%%%%%%%%%%%%%%%%%%%%%%%%%%%%%%%%%%%%%%%%%%%%%%%%%%%%%%%%%%%%
for $\{\textbf{q}\cdot\textbf{R}_t\neq 2\pi N\}$, where $N$ is an integer, gives
%%%%%%%%%%%%%%%%%%%%%%%%%%%%%%%%%%%%%%%%%%%%%%%%%%%%%%%%%%%%%%%%%%%%%%%%%%%%%%%%%%%
\begin{equation}
m_{s\alpha,p}(\textbf{q})=\frac{1}{2}\left(A_{s\alpha,p}-iB_{s\alpha,p}\right).
\label{mqV}
\end{equation}
%%%%%%%%%%%%%%%%%%%%%%%%%%%%%%%%%%%%%%%%%%%%%%%%%%%%%%%%%%%%%%%%%%%%%%%%%%%%%%%%%
Eq.\ (\ref{mqV}) demonstrates that the real and imaginary parts of the eigenvector directly provide the amplitudes in Eq.\ (\ref{EqGen}),
%%%%%%%%%%%%%%%%%%%%%%%%%%%%%%%%%%%%%%%%%%%%%%%%%%%%%%%%%%%%%%%%%%%%%%%%%%%%%%%%%%%
\begin{equation}
\begin{split}
 & \frac{1}{2}A_{s\alpha,p}= m_{0\alpha}\Re[V_{s\alpha,p}], \\
 & \frac{1}{2}B_{s\alpha,p}=-m_{0\alpha}\Im[V_{s\alpha,p}].
\label{EqGen2}
\end{split}
\end{equation}
%%%%%%%%%%%%%%%%%%%%%%%%%%%%%%%%%%%%%%%%%%%%%%%%%%%%%%%%%%%%%%%%%%%%%%%%%%%%%%%%%
In particular, the minimization of the free energy yields $m_{0\alpha}\rightarrow 0$ (i.e.\ $A_{s\alpha,p}\rightarrow 0$ and $B_{s\alpha,p}\rightarrow 0$) when the transition temperature is approached and so the paramagnetic state ($\{\textbf{m}_n\rightarrow\textbf{0}\}$) establishes.
Eq.\ (\ref{EqGen}) can be written in the following form
%%%%%%%%%%%%%%%%%%%%%%%%%%%%%%%%%%%%%%%%%%%%%%%%%%%%%%%%%%%%%%%%%%%%%%%%%%%%%%%%%%%
\begin{equation}
\begin{split}
m_{ts\alpha,p}
=  & \sqrt{A_{s\alpha,p}^2+B_{s\alpha,p}^2}\cos\left[\textbf{q}\cdot\textbf{R}_t-\arctan\left(\frac{B_{s\alpha,p}}{A_{s\alpha,p}}\right)\right] \\
=  & 2m_{0\alpha}\sqrt{\Re[V_{s\alpha,p}]^2+\Im[V_{s\alpha,p}]^2} \\
 &\times\cos\left[\textbf{q}\cdot\textbf{R}_t+\arctan\left(\frac{\Im[V_{s\alpha,p}]}{\Re[V_{s\alpha,p}]}\right)\right].
\label{EqGen4}
\end{split}
\end{equation}
%%%%%%%%%%%%%%%%%%%%%%%%%%%%%%%%%%%%%%%%%%%%%%%%%%%%%%%%%%%%%%%%%%%%%%%%%%%%%%%%%%%
We point out that in the paramagnetic limit the equations above apply independently on each spatial direction, $\alpha=\{x,y,z\}$. This means that only one non-zero spatial projection suffices to satisfy Eq.\ (\ref{EqM}) and so to produce a transition temperature equal to $T_{tr}=u_p(\textbf{q})/3k_\text{B}$ for an eigenvalue $u_p(\textbf{q})$. However, away from the paramagnetic limit, for example at low temperatures and/or under the presence of an applied magnetic field, the minimization of $\mathcal{G}$ gives values of the order parameters maximizing their size at every unit cell ($\{m_n=|\textbf{m}_n|\rightarrow 1\}$ at 0K). This situation typically requires the wave modulation of more than a single non-zero spatial component. The components of the helimagnetic phase considered in appendix \ref{1q} are such that they satisfy the description of the eigenvector in Eq.\ (\ref{EqGen4}) and simultaneously preserve a maximum size of the order parameters throughout the lattice for a given value of $T$.

\section{Expression for the Gibbs free energy of a helimagnetic state}
\label{1q}

In this section we derive an expression for the Gibbs free energy that can be used to describe the ferromagnetic state, and 1q helimagnetic states with an arbitrary value of the wave vector $\textbf{q}$. We exploit the fact that the free energy of a single primitive cell in a 1q state is the same throughout the entire crystal, i.e. it does not depend on the unit-cell lattice index $t$. This allows to obtain a general and simpler form of the Gibbs free energy for 1q states.

We consider that the magnetic field is applied perpendicularly to the basal plane of a helimagnetic state.
Since the following algebra considers the dominant isotropic pairwise terms only, i.e.\ when $\{D_{nn',\gamma}^\text{PM}=0\}$, we can arbitrarily choose the helimagnetic plane as the $xy$-plane, and so set $\textbf{B}=B\hat{\textbf{z}}$. The effect of $\textbf{B}$ is to produce a non-oscillatory ferromagnetic component along the $\hat{\textbf{z}}$-direction, i.e.\ (see Eq.\ (\ref{EqGen}))
%%%%%%%%%%%%%%%%%%%%%%%%%%%%%%%%%%%%%%%%%%%%%%%%%%%%%%%%%%%%%%%%%%%%%%%%%%%%%%%%%%%
\begin{equation}
\begin{split}
 & m_{tsx,p}=A_{sx,p}\cos(\textbf{q}\cdot\textbf{R}_t)+B_{sx,p}\sin(\textbf{q}\cdot\textbf{R}_t), \\
 & m_{tsy,p}=A_{sy,p}\cos(\textbf{q}\cdot\textbf{R}_t)+B_{sy,p}\sin(\textbf{q}\cdot\textbf{R}_t), \\
 & m_{tsz,p}=A_{sz,p}.
\label{EqGenFM}
\end{split}
\end{equation}
%%%%%%%%%%%%%%%%%%%%%%%%%%%%%%%%%%%%%%%%%%%%%%%%%%%%%%%%%%%%%%%%%%%%%%%%%%%%%%%%%%%
Introducing Eq.\ (\ref{EqGenFM}) into Eq.\ (\ref{EQE}) and using both the definition of the lattice Fourier transform and Eq.\ (\ref{util1}), it follows that the internal magnetic energy in the paramagnetic limit is
%%%%%%%%%%%%%%%%%%%%%%%%%%%%%%%%%%%%%%%%%%%%%%%%%%%%%%%%%%%%%%%%%%%%%%%%%%%%%%%%%%%
\begin{equation}
\begin{split}
 & E
= \\
& -\frac{N_c}{2}\sum_{ss'}\sum^{x,y}_{\alpha}
\frac{1}{2}
 J_{ss'}^\text{PM}(\textbf{q})(A_{s\alpha,p}+iB_{s\alpha,p})(A_{s'\alpha,p}-iB_{s'\alpha,p}) \\
 & -\frac{N_c}{2}\sum_{ss'}
J_{ss'}^\text{PM}(\textbf{0})A_{sz,p}A_{s'z,p}.
\label{Gfou}
\end{split}
\end{equation}
%%%%%%%%%%%%%%%%%%%%%%%%%%%%%%%%%%%%%%%%%%%%%%%%%%%%%%%%%%%%%%%%%%%%%%%%%%%%%%%%%%%
The helimagnetic phase in the $xy$-plane that minimizes the free energy and emerges below $T_\text{max}$ corresponds to $\textbf{q}=\textbf{q}_{peak}$, which leads to the following expression for the Gibbs free energy,
%%%%%%%%%%%%%%%%%%%%%%%%%%%%%%%%%%%%%%%%%%%%%%%%%%%%%%%%%%%%%%%%%%%%%%%%%%%%%%%%%%%
\begin{equation}
\begin{split}
& \mathcal{G}
= \\
& -\frac{N_c}{2}\sum_{ss'}\sum^{x,y}_{\alpha}
J_{ss'}^\text{PM}(\textbf{q}_{peak})\frac{1}{2}
(A_{s\alpha}+iB_{s\alpha})
(A_{s'\alpha}-iB_{s'\alpha}) \\
& -\frac{N_c}{2}\sum_{ss'}
J_{ss'}^\text{PM}(\textbf{0})A_{sz}A_{s'z}
-TS_{mag}-N_cB\sum_sA_{sz}\mu_s,
\label{Gfou}
\end{split}
\end{equation}
%%%%%%%%%%%%%%%%%%%%%%%%%%%%%%%%%%%%%%%%%%%%%%%%%%%%%%%%%%%%%%%%%%%%%%%%%%%%%%%%%%%
where we have dropped the index $p$ since it is now set to $p_\text{max}$ everywhere.
Our lattice Fourier transform analysis exploited here follows now by calculating Eq.\ (\ref{EqFT1}) using the isotropic pairwise interactions computed in the paramagnetic state, whose values are shown in Fig.\ \ref{Fig1}(c), for both a helimagnetic phase ($\textbf{q}_{peak}=0.18(1,1,1)\frac{1}{\sqrt{3}}\frac{2\pi}{a}$),
%%%%%%%%%%%%%%%%%%%%%%%%%%%%%%%%%%%%%%%%%%%%%%%%%%%%%%%%%%%%%%%%%%%%%%%%%%%%%%%%%%%
\begin{equation}
J^\text{PM}(\textbf{q}_{peak})=
\begin{pmatrix}
J^{q}_1      & K^{q}        & K^{q}        & K^{q} \\
K^{q*} & J^{q}_2      & L^{q}        & L^{q} \\
K^{q*} & L^{q*} & J^{q}_2      & L^{q} \\
K^{q*} & L^{q*} & L^{q*} & J^{q}_2
\end{pmatrix},
\label{qpeak}
\end{equation}
%%%%%%%%%%%%%%%%%%%%%%%%%%%%%%%%%%%%%%%%%%%%%%%%%%%%%%%%%%%%%%%%%%%%%%%%%%%%%%%%%%%
where $J^{q}_1=6.86\text{ meV}$, $J^{q}_2=11.21\text{ meV}$, $K^{q}=19.49-18.38i\text{ meV}$, and $L^{q}=19.46-18.36i\text{ meV}$, and for the ferromagnetic phase ($\textbf{q}=\textbf{0}$),
%%%%%%%%%%%%%%%%%%%%%%%%%%%%%%%%%%%%%%%%%%%%%%%%%%%%%%%%%%%%%%%%%%%%%%%%%%%%%%%%%%%
\begin{equation}
J^\text{PM}(\textbf{0})=
\begin{pmatrix}
J^\text{FM} & K^\text{FM} & K^\text{FM} & K^\text{FM} \\
K^\text{FM} & J^\text{FM} & K^\text{FM} & K^\text{FM} \\
K^\text{FM} & K^\text{FM} & J^\text{FM} & K^\text{FM} \\
K^\text{FM} & K^\text{FM} & K^\text{FM} & J^\text{FM}
\end{pmatrix},
\label{q0}
\end{equation}
%%%%%%%%%%%%%%%%%%%%%%%%%%%%%%%%%%%%%%%%%%%%%%%%%%%%%%%%%%%%%%%%%%%%%%%%%%%%%%%%%%%
where $J^\text{FM}=11.66\text{ meV}$ and $K^\text{FM}=26.55\text{ meV}$.
In Eq.\ (\ref{qpeak}) the superscript $^{*}$ denotes the complex conjugate. 
We remark that the matrices to be diagonalized at this point are $4\times 4$ of size since the DMI is omitted.
The largest eigenvalues obtained after diagonalizing Eqs.\ (\ref{qpeak}) and (\ref{q0}) are $J_\text{max}(\textbf{q}_{peak})=93$meV and $J_\text{max}(\textbf{0})=91$meV, with normalized eigenvectors $[0.346+0.326i, 0.508, 0.508, 0.508]$ and $[\frac{1}{2}, \frac{1}{2}, \frac{1}{2}, \frac{1}{2}]$, respectively. Whilst the ferromagnetic part of the magnetic phase is described by four parallel components of the order parameter  along $\hat{\textbf{z}}$, the helimagnetism in the $xy$-plane shows a phase shift of $\arctan(0.326/0.346)=43$ degrees between sublattice $s=1$ and the other sublattices. It therefore follows from Eq.\ (\ref{EqGen4}) that
%%%%%%%%%%%%%%%%%%%%%%%%%%%%%%%%%%%%%%%%%%%%%%%%%%%%%%%%%%%%%%%%%%%%%%%%%%%%%%%%%%%
\begin{eqnarray}
\label{simpq}
 & A_{1x}= a_1, B_{1x}=-b_1, \\
 & A_{1y}= b_1, B_{1y}=a_1, \\
 & A_{1z}= a_z
\end{eqnarray}
%%%%%%%%%%%%%%%%%%%%%%%%%%%%%%%%%%%%%%%%%%%%%%%%%%%%%%%%%%%%%%%%%%%%%%%%%%%%%%%%%%%
for the first sublattice ($s=1$), and
%%%%%%%%%%%%%%%%%%%%%%%%%%%%%%%%%%%%%%%%%%%%%%%%%%%%%%%%%%%%%%%%%%%%%%%%%%%%%%%%%%%
\begin{eqnarray}
\label{simpq}
 & A_{sx}=a_2, B_{sx}=0, \\
 & A_{sy}=0,   B_{sy}=a_2, \\
 & A_{sz}=a_z
\end{eqnarray}
%%%%%%%%%%%%%%%%%%%%%%%%%%%%%%%%%%%%%%%%%%%%%%%%%%%%%%%%%%%%%%%%%%%%%%%%%%%%%%%%%%%
for the rest ($s=2,3,4$).
These parameters ($\{a_1,b_1,a_2,a_z\}$) comprise the minimum possible number of coefficients describing the helimagnetic and ferromagnetic components, i.e.\
%%%%%%%%%%%%%%%%%%%%%%%%%%%%%%%%%%%%%%%%%%%%%%%%%%%%%%%%%%%%%%%%%%%%%%%%%%%%%%%%%%%
\begin{equation}
\begin{split}
 m_{t1x} & =a_1\cos(\textbf{q}\cdot\textbf{R}_t)-b_1\sin(\textbf{q}\cdot\textbf{R}_t) \\
& =\sqrt{a_1^2+b_1^2}\cos\left[\textbf{q}\cdot\textbf{R}_t+\arctan\left(\frac{b_1}{a_1}\right)\right] \\
 m_{t1y} & =b_1\cos(\textbf{q}\cdot\textbf{R}_t)+a_1\sin(\textbf{q}\cdot\textbf{R}_t) \\
%& =\sqrt{a_1^2+b_1^2}\cos\left[\textbf{q}\cdot\textbf{R}_t+\arctan\left(\frac{b_1}{a_1}\right)+\frac{\pi}{2}\right] \\
& =\sqrt{a_1^2+b_1^2}\sin\left[\textbf{q}\cdot\textbf{R}_t+\arctan\left(\frac{b_1}{a_1}\right)\right] \\
 m_{t1z} & =a_z 
\label{ms1}
\end{split}
\end{equation}
%%%%%%%%%%%%%%%%%%%%%%%%%%%%%%%%%%%%%%%%%%%%%%%%%%%%%%%%%%%%%%%%%%%%%%%%%%%%%%%%%%%
and
%%%%%%%%%%%%%%%%%%%%%%%%%%%%%%%%%%%%%%%%%%%%%%%%%%%%%%%%%%%%%%%%%%%%%%%%%%%%%%%%%%%
\begin{equation}
\begin{split}
 m_{tsx} & =a_2\cos(\textbf{q}\cdot\textbf{R}_t) \\
 m_{tsy} & =a_2\sin(\textbf{q}\cdot\textbf{R}_t) \\
 m_{tsz} & =a_z,
\label{ms2}
\end{split}
\end{equation}
%%%%%%%%%%%%%%%%%%%%%%%%%%%%%%%%%%%%%%%%%%%%%%%%%%%%%%%%%%%%%%%%%%%%%%%%%%%%%%%%%%%
for $s=1$ and $s>1$, respectively. 
Such an arrangement of coefficients satisfies the eigensolutions obtained for the diagonalization of Eqs.\ (\ref{qpeak}) and (\ref{q0}) and at the same time preserves the total size of the local order parameters from one unit cell to another, i.e.\ $m_{t1}=\sqrt{a_1^2+b_1^2+a_z^2}$ and $m_{ts=2,3,4}=\sqrt{a_2^2+a_z^2}$ do not change as a function of $t$. This allows to write the entropy term as $S_{mag}=N_c(S_1+S_2+S_3+S_4)$ and a convenient expression for the Gibbs free energy per unit cell whose minimization maximizes the size of order parameters at all unit cells for a given value of $T$,
%%%%%%%%%%%%%%%%%%%%%%%%%%%%%%%%%%%%%%%%%%%%%%%%%%%%%%%%%%%%%%%%%%%%%%%%%%%%%%%%%%%
\begin{equation}
\begin{split}
& \frac{1}{N_c}\mathcal{G} 
=
-\frac{1}{2}
J_\text{max}^\text{PM}(\textbf{q}_{peak})
(a_1^2+b_1^2+3a_2^2) -\frac{1}{2}
J_\text{max}^\text{PM}(\textbf{0})4a_z^2 \\
& -T(S_1+S_2+S_3+S_4)-4Ba_z\mu_{Mn} \\
=
& 4\Bigg[-\frac{1}{2}
J_\text{max}^\text{PM}(\textbf{q}_{peak})
a_h^2 -\frac{1}{2}
J_\text{max}^\text{PM}(\textbf{0})a_z^2
-TS-Ba_z\mu_{Mn}\Bigg] \\
%=
%& 4\Bigg[-\frac{1}{2}
%J_\text{max}^{PM}(\textbf{q}_{peak})
%m^2\sin^2\theta -\frac{1}{2}
%J_\text{max}^{PM}(\textbf{0})m^2\cos^2\theta \\
%& -TS-Bm\cos\theta\mu_{Mn}\Bigg],
\label{Gfouu}
\end{split}
\end{equation}
%%%%%%%%%%%%%%%%%%%%%%%%%%%%%%%%%%%%%%%%%%%%%%%%%%%%%%%%%%%%%%%%%%%%%%%%%%%%%%%%%%%
where $\mu_{Mn}$ is the magnitude of the local moment at the Mn site, $S\equiv S_1=S_2=S_3=S_4$, and $a_h^2\equiv a_1^2+b_1^2\approx a_2^2$ has been approximated, which is justified by the components of the eigenvector obtained for Eq.\ (\ref{qpeak}).
The critical line for the 1q state shown in Fig.\ \ref{Fig3}(b) is obtained by minimizing Eq.\ (\ref{Gfouu}) with respect to
$\{a_h, a_z \}$
at different values of $T$ and $B$. $S$, which is unequivocally given by $m_n^2=a_h^2+a_z^2$, is calculated using Eq.\ (\ref{EqS1}) together with Eq.\ (\ref{Eqm}).

\bibliography{./MnGe.bib}

\end{document}